\documentclass[aps,prx,floatfix,twocolumn]{revtex4-1}

\usepackage{amsmath,amssymb}
\usepackage[pdftex]{hyperref,graphicx}
\hypersetup{colorlinks = true, urlcolor = blue, linkcolor = blue, citecolor = blue}

\usepackage{physics}
\usepackage{xcolor}
\usepackage{bm}

\begin{document}
	\title{Band topology, Hubbard model, Heisenberg model, and Dzyaloshinskii-Moriya interaction in twisted bilayer WSe$_2$}

	\author{Haining Pan}
	\affiliation{Condensed Matter Theory Center and Joint Quantum Institute, Department of Physics, University of Maryland, College Park, Maryland 20742, USA}
	\author{Fengcheng Wu}
	\email{wufcheng@umd.edu}
	\affiliation{Condensed Matter Theory Center and Joint Quantum Institute, Department of Physics, University of Maryland, College Park, Maryland 20742, USA}
	\author{Sankar Das Sarma}
	\affiliation{Condensed Matter Theory Center and Joint Quantum Institute, Department of Physics, University of Maryland, College Park, Maryland 20742, USA}

	\begin{abstract}
	We present a theoretical study of single-particle and many-body properties of twisted bilayer WSe$_2$. For single-particle physics, we calculate the band topological phase diagram and electron local density of states (LDOS), which are found to be correlated. By comparing our theoretical LDOS with those measured by scanning tunneling microscopy, we comment on the possible topological nature of  the first moir\'e valence band. For many-body physics, we construct a generalized Hubbard model on a triangular lattice based on the calculated single-particle moir\'e bands. We show that a layer potential difference, arising, for example, from an applied electric field, can drastically change the non-interacting moir\'e bands, tune the spin-orbit coupling in the Hubbard model, control the charge excitation gap of the Mott insulator at half filling, and generate an effective Dzyaloshinskii-Moriya interaction in the effective Heisenberg model for the Mott insulator. Our theoretical results agree with transport experiments on the same system in several key aspects, and establish twisted bilayer WSe$_2$ as a highly tunable system for studying and simulating strongly correlated phenomena in the Hubbard model.    
	\end{abstract}

	\maketitle

	\section{Introduction}\label{sec:intro}
	Twisted bilayers with a long-range moir\'e pattern provide highly tunable platforms to study fundamental physics for both single-particle and many-body phenomena. An important breakthrough was the experimental discovery of superconducting and correlated insulating states \cite{cao2018correlated, cao2018unconventional} in magic-angle twisted bilayer graphene (TBG) \cite{bistritzer2011moire}.  While magic-angle TBG is under active study and hosts a rich variety of phenomena \cite{yankowitz2019tuning,lu2019superconductors,sharpe2019emergent, serlin2020intrinsic}, it poses challenges for both experiment and theory. In experiment, superconducting and correlated insulating states in TBG are fragile and appear only within a narrow range of twist angle around the magic angle ($\sim 1.1^{\circ}$), requiring great experimental efforts to fine tune the twist angle. In theory, the low-energy moir\'e bands in TBG defy the construction of fully symmetric Wannier states because of intrinsic obstructions \cite{po2018origin}, which complicates theoretical analysis.   
	
	It was theoretically proposed that twisted bilayer transition metal dichalcogenides (TMDs) represent a simpler system  compared to TBG and can provide a platform to simulate model Hamiltonians such as Hubbard model and Kane-Mele model \cite{wu2018hubbard, wu2019topological}. Here, TMDs refer to group-VI semiconducting transition metal dichalcogenides such as WSe$_2$ \cite{xiao2012coupled}. The simplicity of TMDs compared to graphene originates from the fact that the former is a semiconductor with a large band gap as well as a large spin-orbit coupling, while the latter is a semimetal with Dirac cones and spin SU(2) symmetry. Because of the reduced symmetries in TMDs, the low-energy degrees of freedom in twisted bilayer TMDs are fewer than TBG, which leads to theoretical simplification, allowing effective realizations of simple yet important model Hamiltonians \cite{wu2018hubbard, wu2019topological}. Another noticeable difference between twisted bilayer TMD and TBG is that the nearly flat moir\'e bands  appear in a large range of twist angles in the former system, but only occur within a small window $(\pm 0.1^{\circ})$ around the magic angle in the latter system. This difference could lead to practical advantages, as there is no longer an acute need to carefully fine tune the twist angle in order to achieve the flat band situation.  Single-particle flat bands strongly enhance the relative interaction strength since the non-interacting kinetic energy is suppressed under the flat-band condition, potentially leading to many interesting correlated quantum phases.
	
	There are two types of twisted TMD bilayers, namely, heterobilayers and homobilayers. In heterobilayers, the two layers are, respectively, two different TMD materials, for example WSe$_2$/MoSe$_2$, which automatically lift the layer degeneracy. This moir\'e system can realize a generalized Hubbard model on a triangular lattice formed by effective moir\'e sites \cite{wu2018hubbard}. Such a Hubbard model simulator based on TMD heterobilayers has recently been experimentally realized in Refs.~\onlinecite{tang2020simulation} and \onlinecite{regan2020mott}, which report evidence for Mott insulators and Wigner crystals.
	
	In this paper we focus on twisted TMD homobilayers, where the two layers are formed from the same material. Because of stronger interlayer coupling, homobilayers can potentially be more interesting as well as more tunable compared to heterobilayers.  Our work is motivated by two experimental studies on twisted bilayer WSe$_2$ (tWSe$_2$), where one experiment is based on  scanning tunneling microscope (STM) \cite{zhang2020flat}, and the other is on transport measurement \cite{wang2019magic}. Both experimental papers \cite{zhang2020flat,wang2019magic} report signatures of narrow moir\'e bands in tWSe$_2$, and the transport experiment \cite{wang2019magic} also identifies half-filled correlated insulators that can be sensitively tuned using an external displacement field. 
    
    The purpose of this work is mainly twofold. First, we study the nature of the low-energy non-interacting moir\'e  bands, including their topological character and their mapping to effective lattice models. We present systematic topological phase diagrams characterized by valley Chern numbers as a function of system parameters. 
    We find that the topology of the first moir\'e valence band is closely connected with the pattern of electron density distribution in moir\'e superlattices. By comparing our theoretical local density of states with those measured by STM \cite{zhang2020flat}, we find that the first moir\'e valence band in tWSe$_2$ is likely to be topologically trivial, and can be described by a one-orbital tight-binding model on a triangular lattice. The tight-binding model combined with Coulomb repulsion leads to the realization of an effective Hubbard model for the corresponding interacting system. Second, we demonstrate the convenient tunability provided by an external out-of-plane displacement field in controlling both single-particle as well as many-body properties of TMD homobilayers. For single-particle physics, we show that $V_z$, a layer potential difference generated by the displacement field,  drastically changes the moir\'e band structure, tunes van Hove singularities, and controls the effective spin-orbit coupling in the tight-binding model. For many-body physics, we predict that $V_z$ generates an effective Dzyaloshinskii-Moriya (DM) interaction in the effective Heisenberg model (the spin model for Mott insulator at half filling) associated with the Hubbard model, and acts as a tunable experimental knob that can turn on and off the corresponding correlated (Mott) insulators at half filling. Our theoretical results are consistent with a recent transport experiment in tWSe$_2$ \cite{wang2019magic}. 
    
    We highlight two specific important predictions of our theory. 
    (1) Even in the parameter space where the first moir\'e valence band is topologically trivial, other moir\'e bands can still be topologically nontrivial. This should motivate transport study on the (topologically nontrivial) second and even third moir\'e valence bands by increasing the hole carrier density.
    (2) The  DM interaction breaks spin SU(2) symmetry down to U(1) symmetry, and leads to in plane spin ordering with vector spin chirality in the $120^{\circ}$ antiferromagnetic ground state of the Heisenberg model on a triangular lattice. This field-tunable DM interaction in the moir\'e system is an interesting phenomenon, which may find applications in spintronics.
	
	The remainder of this paper is organized as follows. 
	In Sec.~\ref{sec:moire}, we present a thorough study of moir\'e band structure in tWSe$_2$ with a focus on the topological character and the electron density distribution in real space.
	In Sec.~\ref{sec:field}, we construct a tight-binding model for the first moir\'e valence band in the topologically trivial regime and in the presence of a finite $V_z$. 
	In Sec.~\ref{sec:hubbard}, we construct a Hubbard model for the first moir\'e band by including Coulomb repulsion. We study the Hubbard model at half filling by mapping it to the corresponding Heisenberg model as well as directly by using a mean-field theory. The effects of $V_z$ as well as an out-of-plane magnetic field on many-body physics are also calculated. In Sec.~\ref{sec:conclusion}, we provide a summary and discuss future research directions.
	
		\begin{figure}[t!]
		\includegraphics[width=1\columnwidth]{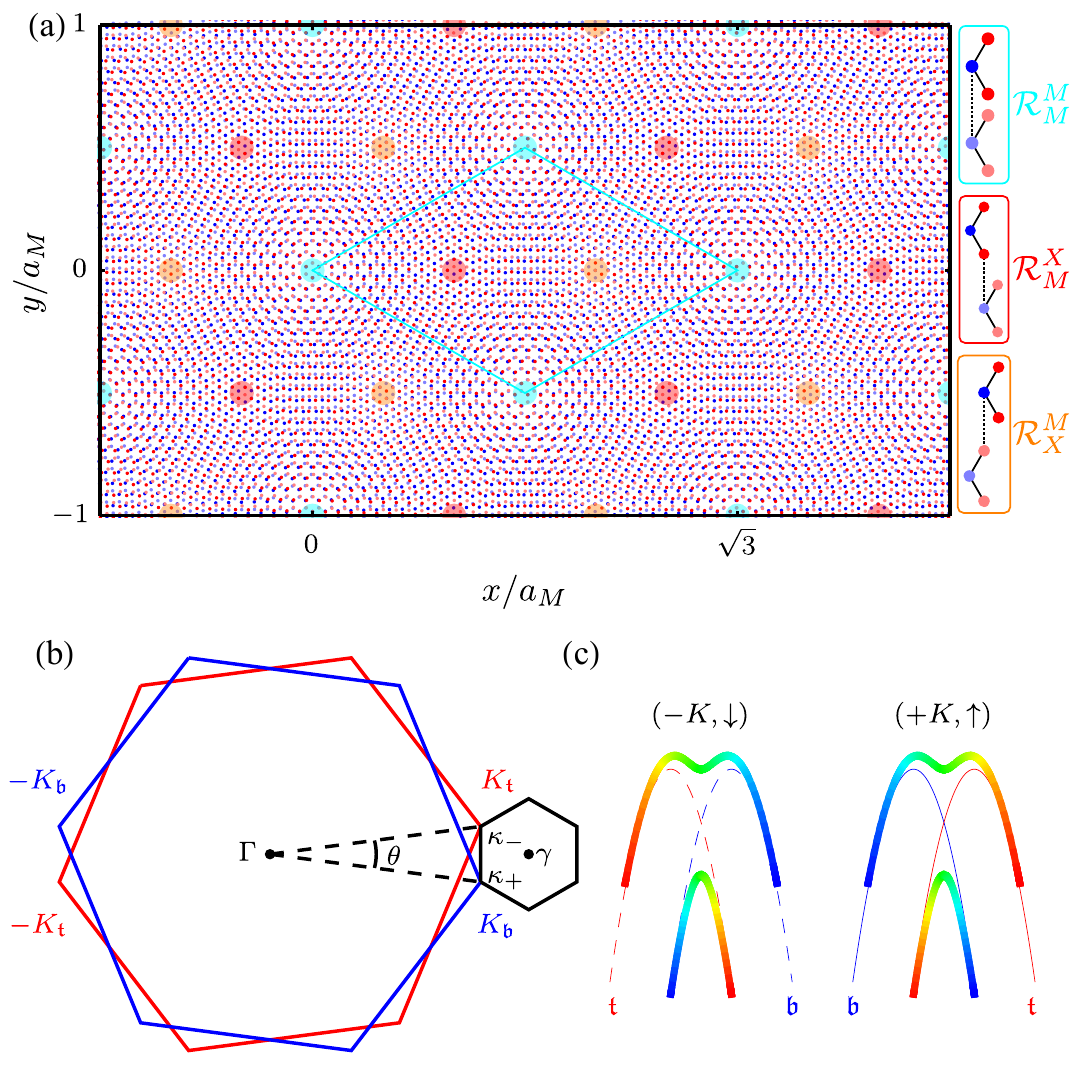}
		\caption{(a) Moir\'e superlattices formed in the twisted bilayer. The dots with cyan, red and orange colors indicate, respectively, $\mathcal{R}_{M}^{M}$, $\mathcal{R}_{M}^{X}$ and $\mathcal{R}_{X}^{M}$ positions  with local stacking configurations shown in the insets. The cyan lines mark a moir\'e unit cell. (b) Brillouin zones associated with the bottom (blue) and top (red) layers, and  the moir\'e Brillouin zone (black). (c) Schematic illustration of band structure in the twisted bilayer.}	
		\label{fig:moire_pattern}
	\end{figure}
	
	\section{moir\'e band structure}
	\label{sec:moire}
	
	\subsection{Moir\'e Hamiltonian}
	
	Twisted TMD homobilayers  with a long-range moir\'e period has two distinct stacking configurations \cite{wu2019topological}, of which the twist angle $\theta$ between the two layers are, respectively, near $0^{\circ}$ and $180^{\circ}$. These two configurations are different because each monolayer TMD has a $D_{3h}$ point-group symmetry without $C_{2z}$ symmetry (i.e., twofold rotation around out-of-plane $\hat{z}$ axis). The twisted bilayer with $\theta$ close to $180^{\circ}$ can realize a two-orbital Hubbard model on a triangular lattice (see Supplemental Material in Ref.~\onlinecite{wu2019topological}). 
	
	In this work, we focus on valence band states in tWSe$_2$ with a small twist angle $\theta$ near $0^{\circ}$, motivated by recent experimental studies \cite{wang2019magic,zhang2020flat}. This situation has been studied in Ref.~\onlinecite{wu2019topological} for the single-particle moir\'e bands. Here we present a more systematic investigation including a complete topological phase diagram and a microscopic many-body theory. As shown in Fig.~\ref{fig:moire_pattern}(a), the  moir\'e pattern formed in the twisted bilayer has a period $a_M \approx a_0/|\theta|$, where $a_0 \approx 3.28 \, \text{\AA}$  is the monolayer lattice constant. In each moir\'e unit cell (MUC), there are three high-symmetry positions: $\mathcal{R}_{M}^M$, $\mathcal{R}_{M}^X$ and $\mathcal{R}_{X}^M$, where  $M$ and $X$, respectively, represent metal and chalcogen atoms, and $\mathcal{R}_{\alpha}^{\beta}$ marks a local position where the $\alpha$ atom in the bottom layer is vertically aligned with the $\beta$ atom in the top layer. The twisted bilayer has $D_3$ point-group symmetry generated by a threefold rotation $C_{3z}$ around the  $\hat{z}$ axis and
	a twofold rotation $C_{2y}$ around the in-plane $\hat{y}$ axis that swaps the two layers. The $D_3$ point group is reduced to $C_3$ when an external out-of-plane displacement field is applied to the system.
	
	In semiconducting TMDs, the topmost valence band states at $\pm K$ valleys and $\Gamma$ valley can be close in energy \cite{liu2013threeband}. For small angle tWSe$_2$, STM measurement shows that its topmost moir\'e valence bands originate from $\pm K$ valleys instead of  $\Gamma$ valley\cite{zhang2020flat}. Therefore, we focus on $\pm K$ valleys states.

	\begin{figure}[t]
		\centering
		\includegraphics[width=1\columnwidth]{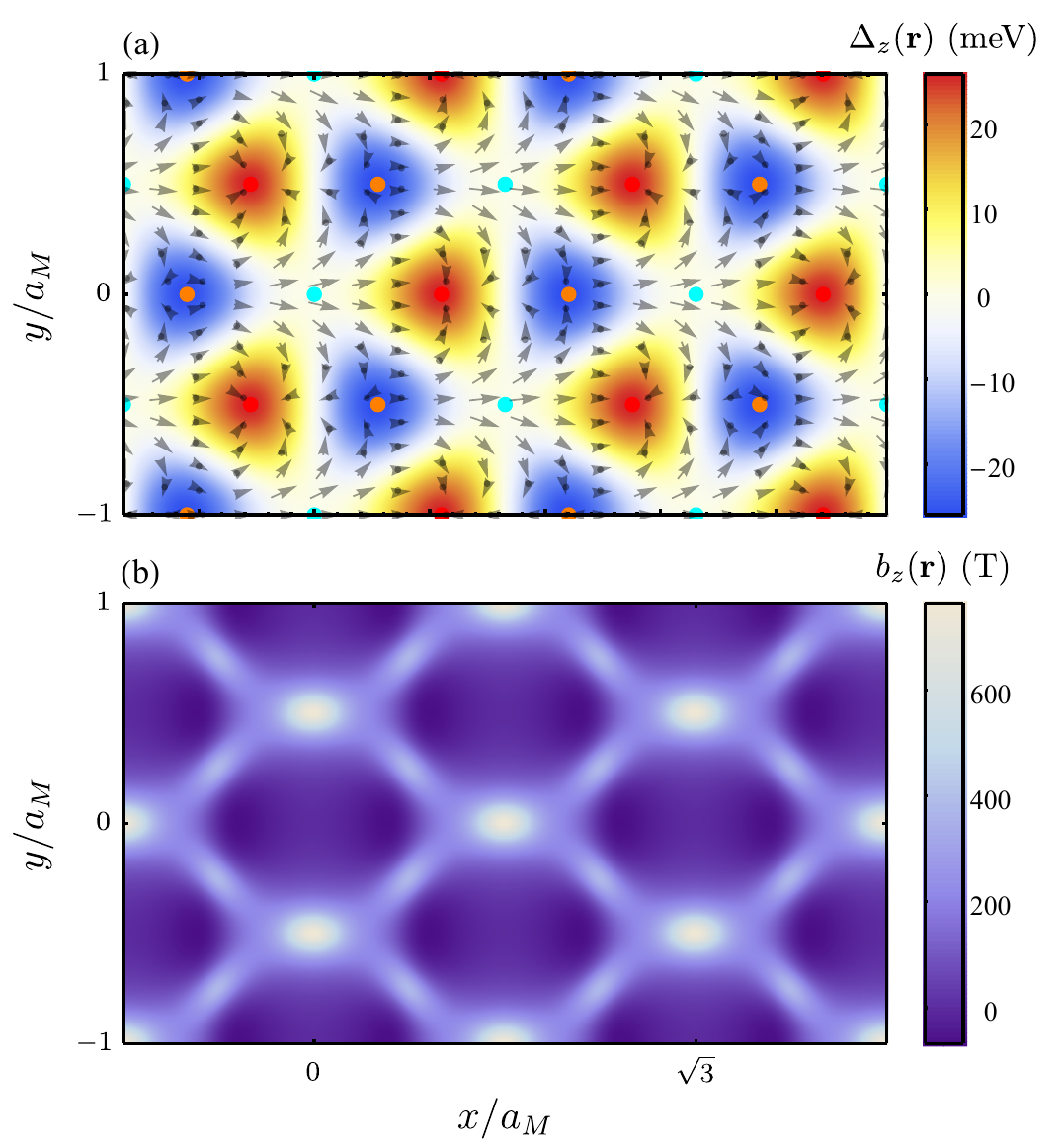}
		\caption{(a) The spatial variation of the layer pseudospin magnetic field $\bm{\Delta}(\bm{r})$ in the moir\'e pattern. The arrows represent the $x$ and $y$ components of $ \bm{\Delta}(\bm{r}) $ and the color map shows the $ z $  component. The cyan, red and orange dots mark high-symmetry positions as in Fig.~\ref{fig:moire_pattern}(a). (b) The effective magnetic field $ b_z(\bm{r}) $ that corresponds to the skyrmion field in (a).  Parameter values are $\left(\theta,V,\psi,w\right)=(3^\circ,5\text{ meV}, 0.5\pi, 20\text{ meV})$.}	
		\label{fig:skyrmion}
	\end{figure}

    \begin{figure}[t!]
		\centering
		\includegraphics[width=1\columnwidth]{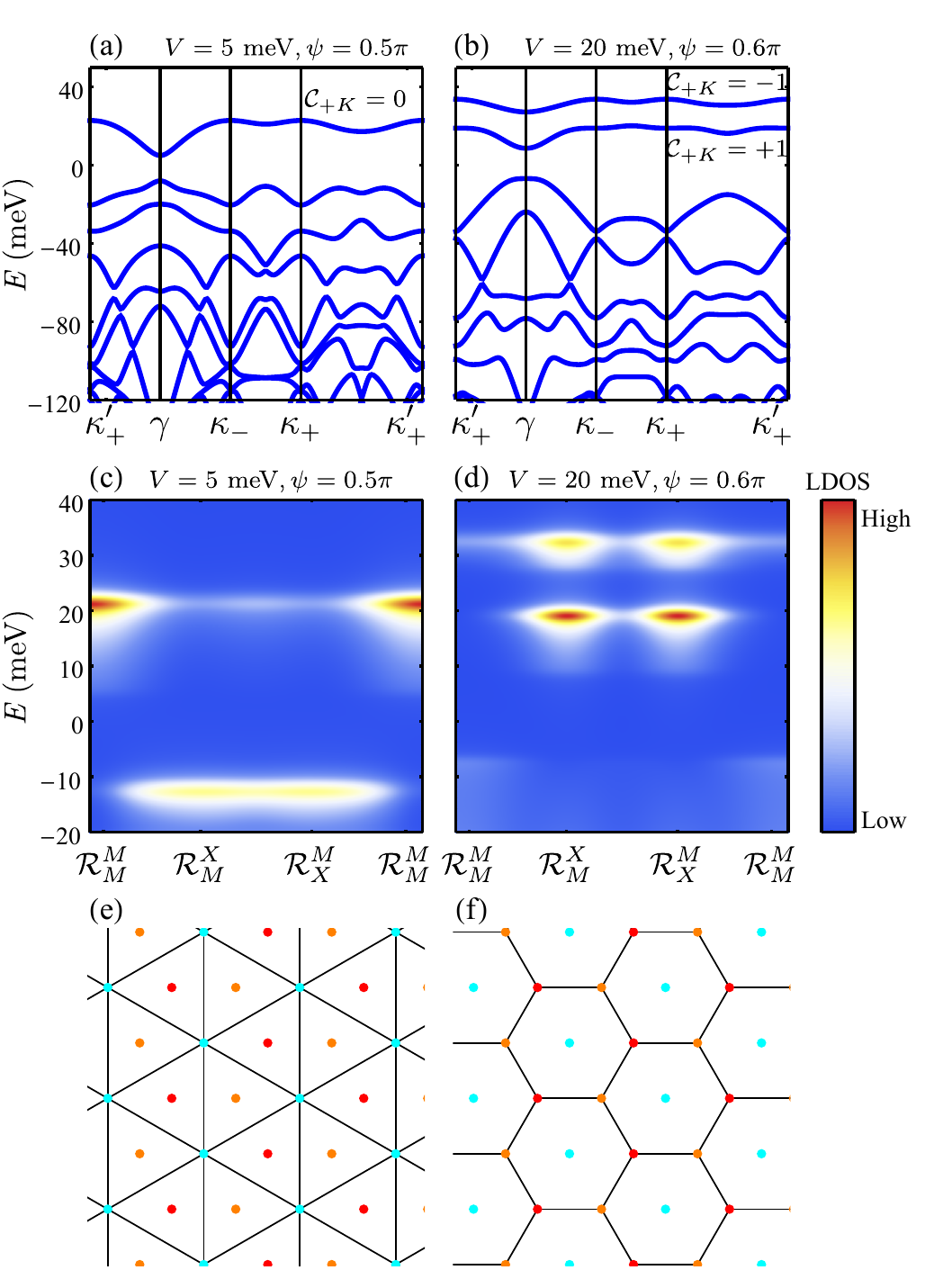}
		\caption{(a) and (b) moir\'e band structure for different values of $(V, \psi)$. $\theta$ is $3^{\circ}$ and $w$ is 20 meV. (c) Local density of states (LDOS) for the moir\'e bands in (a). The horizontal axis is along a high symmetry line in the moir\'e pattern. (d) Similar as (c) but for moir\'e bands in (b). (e) An effective triangular lattice model for the first moir\'e band in (a). (f) An effective honeycomb lattice model for the first and second moir\'e bands in (b). In (e) and (f), the cyan, red and orange dots mark $\mathcal{R}_{M}^{M}$, $\mathcal{R}_{M}^{X}$ and $\mathcal{R}_{X}^{M}$ positions in the moir\'e pattern.}	
		\label{fig:moire_band}
    \end{figure}		
    
        \begin{figure}[t]
		\centering
		\includegraphics[width=0.8\columnwidth]{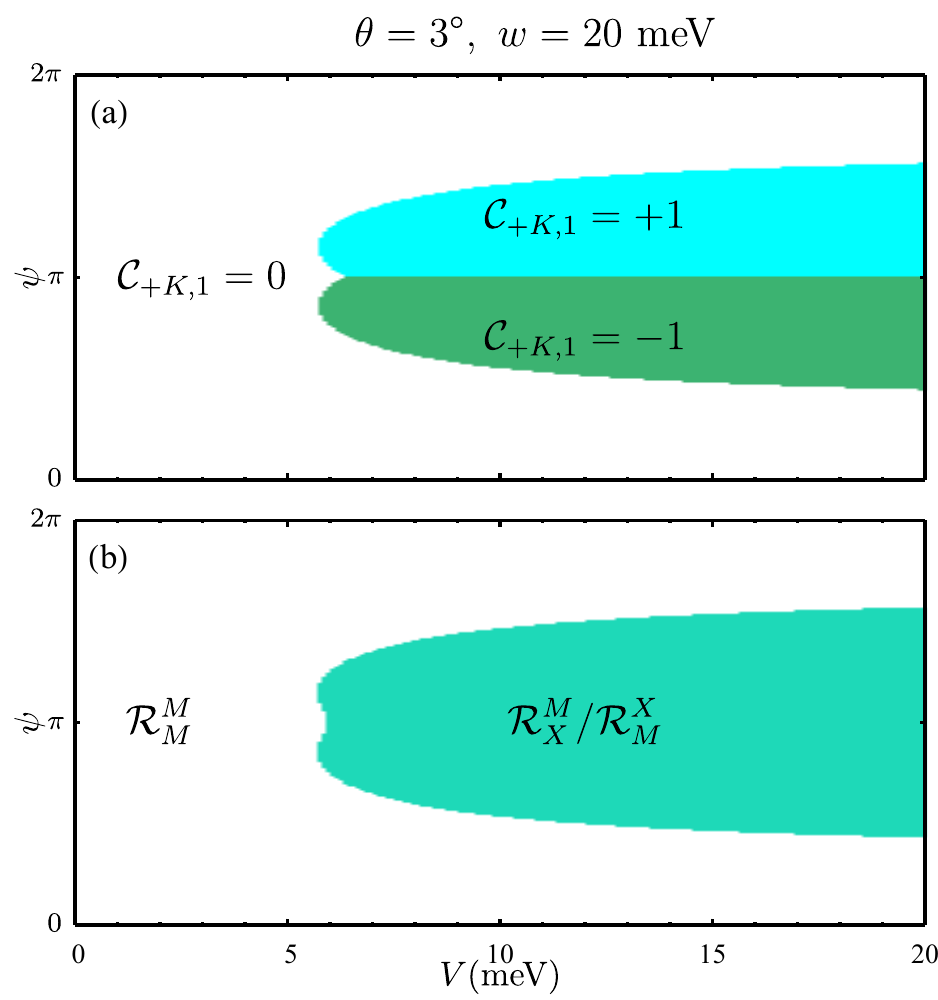}
		\caption{(a) Topological phase diagram characterized by the Chern number $C_{+K,1}$ of the first moir\'e valence band. (b)The white (green) regime represents parameter space where the potential maximum positions of $\tilde{\Delta}$ are at $\mathcal{R}_M^M$ ($\mathcal{R}_M^X$/$\mathcal{R}_X^M$). The phase boundary between topological ($C_{+K,1} \neq 0$) and trivial ($C_{+K,1} = 0$) phases in (a) closely follows the boundary between white and green regimes in (b).  }	
		\label{fig:phase_diagram}
    \end{figure}

    \begin{figure}[t]
		\centering
		\includegraphics[width=0.95\columnwidth]{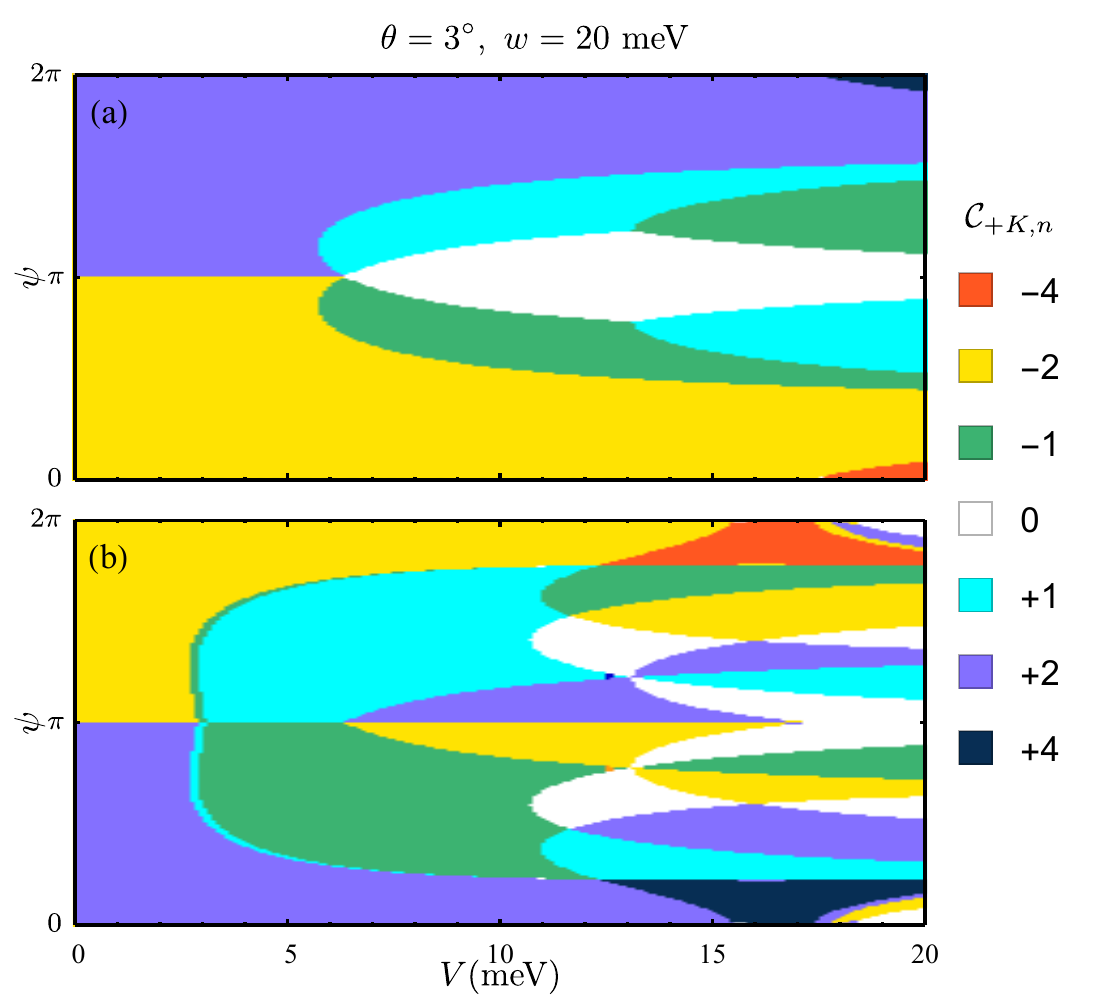}
		\caption{(a) Topological phase diagram characterized by the Chern number $C_{+K,2}$ of the second moir\'e valence band. The colors encode different integer values of $C_{+K,2}$. (b) Similar as (a) but for the Chern number of the third moir\'e valence band.  }	
		\label{fig:phase_diagram_23}
    \end{figure}
	
	There is a large valley-dependent spin splitting in the valence bands at $\pm K$ valley, which leads to an effective spin-valley locking \cite{xiao2012coupled} and reduces the degrees of freedom in the low-energy theory. Therefore, we only consider the spin up (down) valence band in $+K$($-K$) valley, as schematically shown in Fig.~\ref{fig:moire_pattern}(c).  Furthermore, we treat $+K$ and $-K$ valleys separately in the single-particle Hamiltonian because the two valleys are separated by a large momentum when $\theta$ is small [Fig.~\ref{fig:moire_pattern}(b)]. Since the two valleys are related by time-reversal symmetry $\mathcal{T}$, we can focus on $+K$ valley, of which the moir\'e Hamiltonian is given by \cite{wu2019topological}
	\begin{equation}
	\mathcal{H}_\uparrow=\begin{pmatrix}
	-\frac{\hbar^2 (\bm{k}-\bm{\kappa}_+)^2}{2m^*}+\Delta_{+}(\bm{r}) & \Delta_{\text{T}}(\bm{r})\\
	\Delta_{\text{T}}^\dagger(\bm{r}) & -\frac{\hbar^2(\bm{k}-\bm{\kappa}_-)^2}{2m^*}+\Delta_{-}(\bm{r})
	\end{pmatrix},
	\label{eq:moire}
	\end{equation}
	where the $2\times 2$ matrix is in the layer pseudospin space, the diagonal terms are associated with each layer, and the off-diagonal terms describe the interlayer tunneling. In Eq.\eqref{eq:moire}, $m^*$ is the valence band effective mass, the layer-dependent momentum offset $\bm{\kappa}_{\pm}=[4\pi/(3a_M)](-\sqrt{3}/2,\mp 1/2)$ capture the rotation in the momentum space [Fig.~\ref{fig:moire_pattern}(b)], and $\Delta_{\pm}(\bm{r})$ is the layer-dependent moir\'e potential given by
	\begin{equation}
	\Delta_{\pm}(\bm{r}) = 2 V \sum_{j=1,3,5}^{}\cos(\bm{b}_j\cdot \bm{r} \pm \psi),
	\end{equation}
	where $V$ and $\psi$ respectively characterize the amplitude and spatial pattern of the moir\'e potential, and $\bm{b}_j$ is the moir\'e reciprocal lattice vectors in the first shell. Here $\bm{b}_1=[4\pi/(\sqrt{3}a_M)](1,0)$ and $\bm{b}_j$ with $j=2,3\dots6$ are related to $\bm{b}_1$ by $(j-1)\pi/3$ rotation. 
	The interlayer tunneling $\Delta_{\text{T}}(\bm{r})$ is parametrized by
	\begin{equation}
	\Delta_{\text{T}}(\bm{r}) = w (1+e^{-i \bm{b}_2 \cdot \bm{r}}+e^{-i \bm{b}_3 \cdot \bm{r}}),
	\end{equation}
	where $w$ is the interlayer tunneling strength.
	
	We take the effective mass $m^*$ to be 0.45$m_0$ following the experimental value \cite{fallahazad2016shubnikov} of monolayer WSe$_2$, where $m_0$ is the electron rest mass. Other parameters $(V, \psi, w)$ could in principle be  estimated using first-principles  calculations \cite{wu2018hubbard, wu2019topological, naik2018ultraflatbands, zhang2019moir}. However, such estimations may suffer from large uncertainties as these parameters are very sensitive to the layer separation that  varies spatially in the moir\'e pattern. Therefore, we treat $(V, \psi, w)$ as phenomenological parameters, and present a systematic study of the moir\'e band structure as a function of these parameters. At this early stage of the development of the subject, first-principles band structure calculations, with their inherent quantitative uncertainties, should be used with caution in developing low-energy effective theories with small energy scales, where the relevant band parameters can be obtained from experimental measurements (or can be taken as unknown phenomenological parameters of the effective theory).

	\subsection{Layer pseudospin skyrmion}
	From the continuum Hamiltonian $\mathcal{H}_{\uparrow}$ in the layer pseudospin space, we can define a scalar potential $\Delta_0$ and a layer pseudospin magnetic field $\bm{\Delta}$ as follows:
	\begin{equation} \label{key}
	\begin{aligned}
	\Delta_0(\bm{r})&=\frac{\Delta_+ + \Delta_-}{2},\\
	\bm{\Delta}(\bm{r})&=\left(\text{Re}\Delta_{\text{T}}^\dagger,\text{Im}\Delta_{\text{T}}^\dagger, \frac{\Delta_+ - \Delta_-}{2} \right).
	\end{aligned}
	\end{equation}
	We plot the layer pseudospin magnetic field $\bm{\Delta}$ in Fig.~\ref{fig:skyrmion}(a). The in-plane vector $(\Delta_x, \Delta_y)$,  which accounts for interlayer tunneling, forms vortices and antivortices  around $\mathcal{R}_{M}^X$ and $\mathcal{R}_{X}^M$ positions, while $\Delta_z$, the $z$ component of $\bm{\Delta}$, takes maximum and minimum values at these two high-symmetry positions.  This spatial profile indicates that $\bm{\Delta}$ forms a skyrmion lattice, which is characterized by the following winding number $N_w$~\cite{nagaosa2013topological}:
	\begin{equation}\label{eq:windingnumber}
	\begin{aligned}
	N_w &\equiv\frac{1}{4\pi}\int_{\text{MUC}}d\bm{r}\frac{\bm{\Delta}\cdot(\partial_x \bm{\Delta} \times \partial_y \bm{\Delta})}{\abs{\bm{\Delta}}^3}\\
	&=\begin{cases}
	+1, V\sin\psi>0\\
	-1, V\sin\psi<0
	\end{cases}.
	\end{aligned}
	\end{equation}
	Here $N_w$ is quantized to $+1$ or $-1$ depending on the sign of $V \sin \psi$. 
	
	In the adiabatic limit where the electron's pseudospin  follows the skyrmion texture locally, electron's wave function acquires a real-space Berry phase \cite{nagaosa2013topological}, which can be attributed to an emergent (fictitious) orbital magnetic field $b_z$ which is pointing out of plane:  
	\begin{equation}\label{eq:bz}
	b_z(\bm{r})=\frac{\hbar}{2e}\frac{\bm{\Delta}\cdot(\partial_x \bm{\Delta} \times \partial_y \bm{\Delta})}{\abs{\bm{\Delta}}^3}.
	\end{equation}
	The effective magnetic flux produced by $b_z$ over one MUC is quantized to $\pm h/e$, following Eq.~\eqref{eq:windingnumber}.
	Figure~\ref{fig:skyrmion}(b) plots the spatial variation of $b_z$ in the moir\'e pattern, and shows that $b_z$ has a strong spatial variation with a large peak value on the order of a few hundreds of teslas, much higher than any real available laboratory magnetic fields.

	The skyrmion lattice and the emergent $b_z$ field open up the possibility for topological moir\'e bands. However, we note that the adiabatic limit is {\it not} always satisfied in our system, and we find that the skyrmion winding number and the band topology do not have a one-to-one correspondence.
	
	We also define an effective total potential $\tilde{\Delta}=\Delta_0+|\bm{\Delta}|$. Because the kinetic energy in Eq.~\eqref{eq:moire} has a hole-type dispersion, low-energy states in our theory are those that are close to the valence band edge. In a semiclassical picture, low-energy states near the band edge tend to be confined near positions where $\tilde{\Delta}$ reaches its maximum value. The maximum positions of $\tilde{\Delta}$ can be at $\mathcal{R}_{M}^{M}$ or  $\mathcal{R}_{M}^{X}$/$\mathcal{R}_{X}^{M}$ depending on the exact values of $(V,\psi, w)$, which can have important implications on the band topology, as discussed in the following.

	\begin{figure}[t]
	\centering
	\includegraphics[width=1\columnwidth]{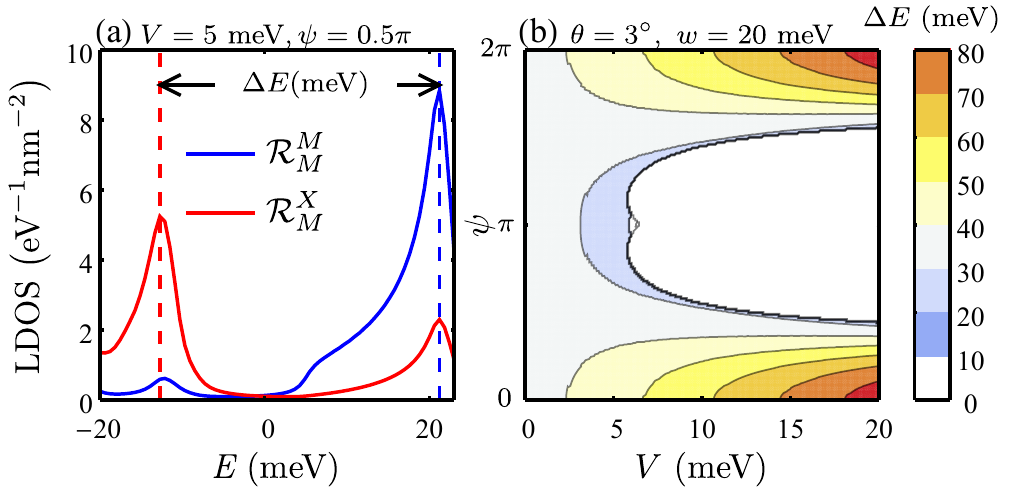}
	\caption{(a) Local density of states at $\mathcal{R}_M^M$ (blue curve) and $\mathcal{R}_M^X$ (red curve) positions. The first (second) peak marked by blue (red) dashed line is mainly at $\mathcal{R}_M^M$  ($\mathcal{R}_M^X$) positions. Parameter values are the same as those for Fig.~\ref{fig:moire_band}(a). We use $\Delta E$ to denote the energy separation between the first and second peaks. (b) $\Delta E$ as a function of $V$ and $\psi$ with $w = 20$ meV and $\theta =3^{\circ}$. The color map shows the value of $\Delta E$ in the topologically trivial regime of Fig.~\ref{fig:phase_diagram}(a). The white region corresponds to the topological phases of Fig.~\ref{fig:phase_diagram}(a), where we do not present the value of $\Delta E$.}	
	\label{fig:LDOS}
    \end{figure}	
    
	\begin{figure*}[t]
		\centering
		\includegraphics[width=2\columnwidth]{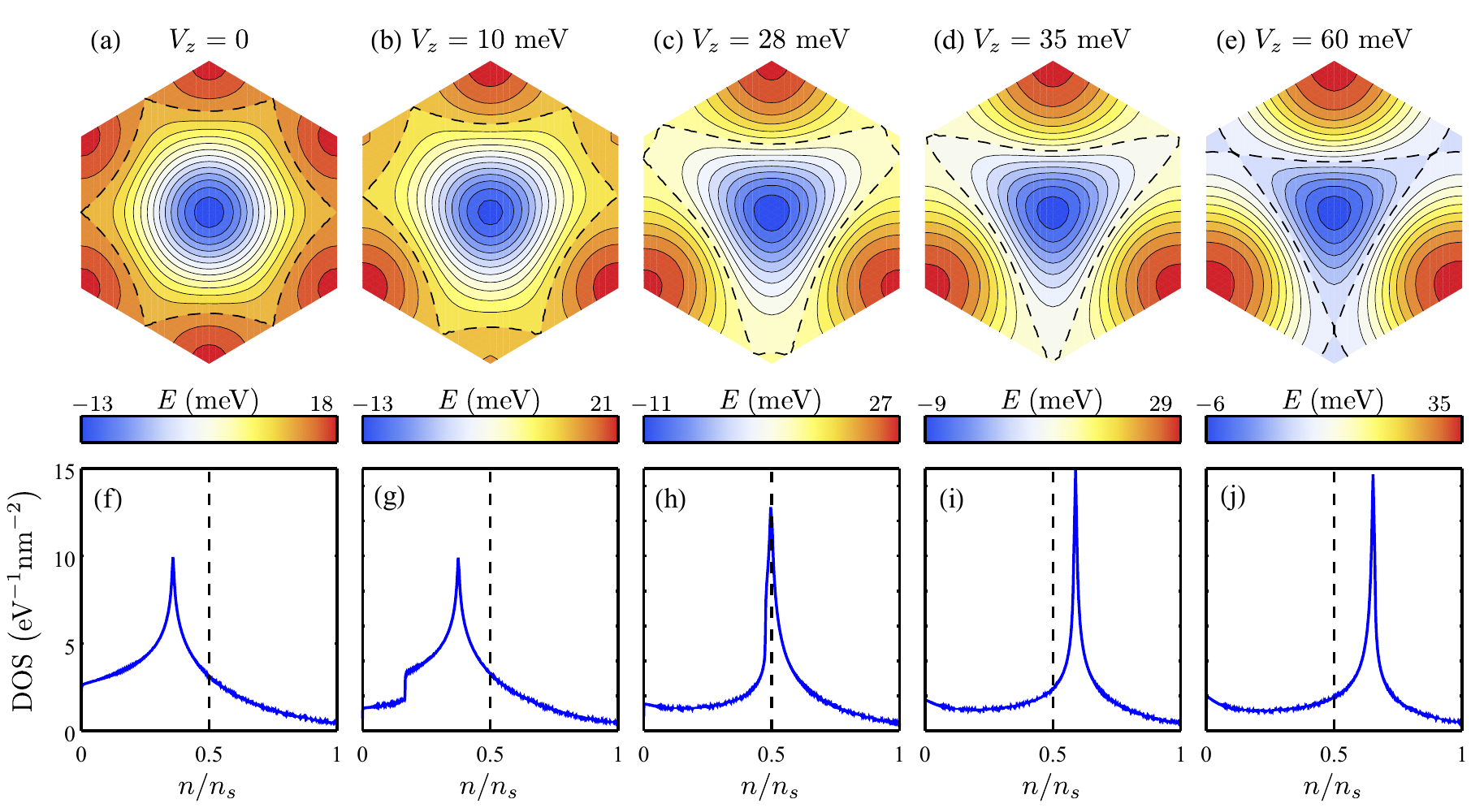}
		\caption{(a)-(e) The first moir\'e valence band in $+K$ valley for different values of $V_z$. The dashed lines mark the Fermi contour at the van Hove energy. Parameter values are $(\theta, V, \psi, w) = (4^{\circ}, 4.4 \text{meV}, 5.9, 20 \text{meV})$. The plotted band is topologically trivial.  (f)-(j) The corresponding density of states for the band shown in (a)-(e). The horizontal axis represents the hole filling factor $n/n_s$. The first moir\'e valence bands are fully filled (empty) at $n/n_s=$0 (1).}	
		\label{fig:2D_band}
	\end{figure*}

	\subsection{Topological phase diagram}
	
	We diagonalize the moir\'e Hamiltonian in Eq.~\eqref{eq:moire} using plane-wave expansion based on Bloch's theorem, and show representative moir\'e band structure in Fig.~\ref{fig:moire_band}. To discuss band topology, we use $\mathcal{C}_{\pm K, n}$ to denote the Chern number of the $n$-th moir\'e valence band in $\pm K$ valleys. Here, we label the moir\'e valence bands in a descending order of energy, and the topmost moir\'e valence band in each valley is labeled as the first one. We focus our discussion on $+K$ valley, since $\mathcal{C}_{-K, n} = -\mathcal{C}_{+K, n}$ because of time-reversal symmetry. 
	
	We find that the topological character of the moir\'e bands depends on the precise values of the band parameters. In Fig.~\ref{fig:moire_band}(a), the first moir\'e band is topologically trivial with a zero Chern number. By contrast, in Fig.~\ref{fig:moire_band}(b) with a different set of parameter values, the first moir\'e band is topologically nontrivial with a finite Chern number. The fact that the topology of the moire bands depends on the details of the parameter values is not surprising since the relevant band Chern number depends on the details of the wave function and is not determined uniquely by any symmetry. For the two sets of parameter values used, respectively, in Figs.~\ref{fig:moire_band}(a) and ~\ref{fig:moire_band}(b), the corresponding skyrmion winding numbers $N_w$ are both quantized to $+1$, which shows that the moir\'e band topology is {\it not} uniquely determined by $N_w$ as the adiabatic limit is {\it not} always satisfied.
	
	The band topology turns out to have a close connection with the spatial pattern of the effective total potential $\tilde{\Delta}$. For Fig.~\ref{fig:moire_band}(a), the corresponding $\tilde{\Delta}$ reaches its potential maximum at $\mathcal{R}_M^M$ positions. Therefore, electrons in the first moir\'e band of Fig.~\ref{fig:moire_band}(a) are confined to $\mathcal{R}_M^M$ positions, which is verified by the local density of states (LDOS) plotted in Fig.~\ref{fig:moire_band}(c).  It follows that the first band in Fig.~\ref{fig:moire_band}(a)  can be described using a tight-binding model on a triangular lattice formed  by $\mathcal{R}_M^M$ sites [Fig.~\ref{fig:moire_band}(e)]. 
	
	As a comparison, the first and second moir\'e bands in Fig.~\ref{fig:moire_band}(b) are topological with Chern numbers of $-1$ and $+1$, respectively. Electron density in both bands is peaked near $\mathcal{R}_M^X$ and $\mathcal{R}_X^M$ positions [Fig.~\ref{fig:moire_band}(d)], following the potential maximum positions of the corresponding $\tilde{\Delta}$. As shown in Ref.~\onlinecite{wu2019topological}, these two topological bands with opposite Chern numbers as a whole can be described by the Haldane model \cite{haldane1988model} on a honeycomb lattice formed by $\mathcal{R}_M^X$ and $\mathcal{R}_X^M$ sites [Fig.~\ref{fig:moire_band}(f)]. Therefore, the full system that consists of $\pm K$ valleys can realize the Kane-Mele model \cite{kane2005quantum} that includes two time-reversed partner copies of the Haldane model.
	
	To obtain a systematical characterization of the band topology, we present a phase diagram in Fig.~\ref{fig:phase_diagram}(a) which plots the Chern number $\mathcal{C}_{+K, 1}$ of the first moir\'e valence band as a function of $V$ and $\psi$ for a fixed value of $w$. There are three phases: the topological phases with $ \mathcal{C}_{+K,1}=  +1$  or $-1$,  and the trivial phase with $ \mathcal{C}_{+K,1}=0 $. The trivial (topological) regime closely tracks the parameter space where the potential maximum positions of $\tilde{\Delta}$ are at $\mathcal{R}_M^M$ ($\mathcal{R}_M^X$/$\mathcal{R}_X^M$), as shown by Figs.~\ref{fig:phase_diagram}(a) and ~\ref{fig:phase_diagram}(b).
	
	The topological phase diagram for the second and third moir\'e valence bands are, respectively, plotted in Figs.~\ref{fig:phase_diagram_23}(a) and ~\ref{fig:phase_diagram_23}(b). The Chern number $C_{+K, n}$ for $n=2$ and $3$ has a complicated dependence on the model parameters, which gives rise to the colorful phase diagrams in Fig.~\ref{fig:phase_diagram_23}. By comparing Figs.~\ref{fig:phase_diagram}(a) and \ref{fig:phase_diagram_23}, we can conclude that the second and third moir\'e bands can be topological even in the parameter space where the first moir\'e valence band is topologically trivial. This has important experimental consequences since, in principle, these higher topological moir\'e bands can be studied experimentally if the chemical potential resides in the higher bands.
		
	\subsection{Comparison with STM experiment}
	
	To determine which phase has actually been realized in tWSe$_2$, we now turn to STM experiments on this system very recently reported in Ref.~\onlinecite{zhang2020flat}. In this experiment \cite{zhang2020flat}, the first LDOS peak at the valence band side (i.e., holes) is found to be primarily localized at $ R_M^M $ positions and the second peak is localized at $R_M^X $ and $ R_X^M $ positions [Fig. 2d in Ref.~\onlinecite{zhang2020flat}], which is consistent with the LDOS structure [Fig.~\ref{fig:LDOS}(a)] in the trivial regime of Fig.~\ref{fig:phase_diagram}(a). With this comparison between the experiment~\cite{zhang2020flat} and our theory, we find that the first moir\'e valence bands in tWSe$_2$ are likely topologically trivial. 
	
	The energy separation $\Delta E$ between the first and second LDOS peaks is found to be $\sim 40$ meV for tWSe$_2$ with $\theta\approx 3^{\circ}$ in Ref.~\onlinecite{zhang2020flat}.  We plot our theoretical value of $\Delta E$ as a function of $V$ and $\psi$ at a fixed value of $w$ in Fig.~\ref{fig:LDOS}(b). The experimental value $\Delta E \approx 40 $ meV constraints $(V,\psi, w)$ to a finite parameter space that belongs to the topologically trivial regime of Fig.~\ref{fig:phase_diagram}(a), but does not lead to a unique determination of $(V,\psi, w)$. We choose a typical set of parameters $ (V,\psi,w)= $(4.4 meV, 5.9, 20 meV), which reproduces the experimental LDOS structure both qualitatively and quantitatively, and use them in all the following calculations.
	
	We make two additional remarks. (1) The experimental LDOS peak energies are subjected to uncertainties, because the experimental LDOS curves are currently broad in energy \cite{zhang2020flat}. Future STM measurement with high resolution is required to fully determine the moir\'e band energetics and local density distribution.  (2) Lattice relaxation effects, which we do not study explicitly in this work, can become important for small twist angles ($\theta < 2.5^{\circ}$) \cite{enaldiev2020stacking}. Therefore, we restrict our study mainly to $\theta \ge 3^{\circ}$.

	\begin{figure}[t]
		\centering
		\includegraphics[width=1\columnwidth]{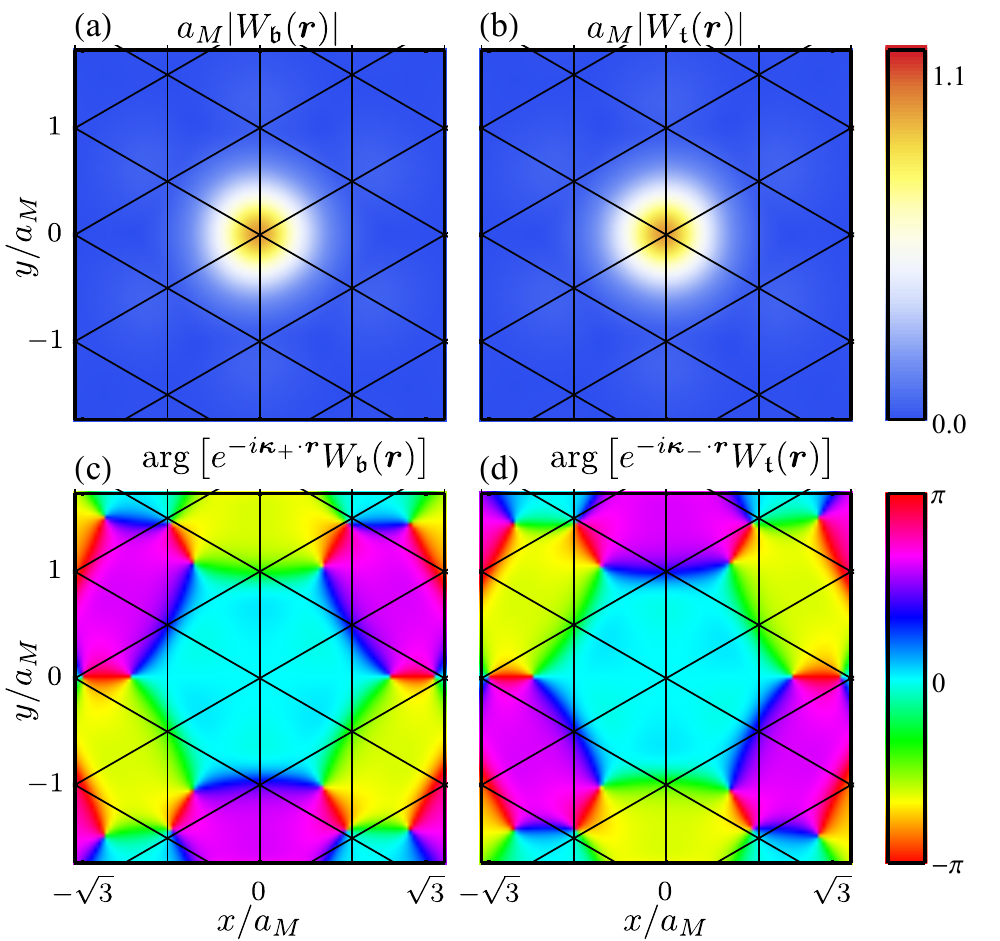}
		\caption{ (a),(b) Amplitude of $W_b(\bm{r})$ and $W_t(\bm{r})$, which are, respectively, the bottom and top layer components of the Wannier state. (c),(d) Phase of $W_\mathfrak{b}(\bm{r})\exp(-i \bm{\kappa}_+ \cdot \bm{r})$ and $W_\mathfrak{t}(\bm{r})\exp(-i \bm{\kappa}_- \cdot \bm{r})$, where the additional phase factors $\exp(-i \bm{\kappa}_{\pm} \cdot \bm{r})$ make the three-fold rotational symmetry transparent. The black lines mark the effective triangular lattice. Parameter values are the same as those used for Fig.~\ref{fig:2D_band}(a).}
		\label{fig:wannier}
	\end{figure}
	
	\section{Field-tunable lattice model}\label{sec:field}
	We focus on the first moir\'e valence band in the topologically trivial regime, and construct an effective tight-binding model for this band in the presence of a layer potential difference $V_z$. We note that $V_z$ can also drive moir\'e bands that are initially in the topological phase to become topologically trivial \cite{wu2019topological}. With the experimentally tunable parameter $V_z$, topologically trivial moir\'e bands can always be realized in tWSe$_2$. 
	
	The potential $V_z$ is generated by an external out-of-plane displacement field, and is a tuning knob in controlling the band structure as well as many-body physics.  With a finite $V_z$, we replace $\Delta_{\pm}$ in the moir\'e Hamiltonian of Eq.~(\ref{eq:moire})  by  $\Delta_{\pm} \pm V_z/2$. At $V_z=0$, the wave function of the first moir\'e band in $+K$ valley at the two corners of the moir\'e Brillouin zone, $\kappa_+$ and $\kappa_-$, are primarily located in the bottom and top layers, respectively. A finite layer potential difference $V_z$ shifts the band energies at $\kappa_+$ and $\kappa_-$ in opposite ways, and therefore, can lead to a drastic change in the band structure as demonstrated in Figs.~\ref{fig:2D_band} (a)-(e). A noticeable effect is that the van Hove saddle points in the band structure can be effectively moved in the moir\'e Brillouin zone by tuning $V_z$. There is a critical value of $V_{z}$, at which three van Hove saddle points merge to a single higher-order saddle point \cite{yuan2019magic,bi2019excitonic,wu2019ferromagnetism} at one of the corners of the moir\'e Brillouin zone [Fig.~\ref{fig:2D_band}(d)]. As a result, the van Hove singularities in the density of states (DOS) can be tuned from below to above half-filling by changing $V_z$ as shown in  Figs.~\ref{fig:2D_band} (f)-(j). This can have important implications on many-body physics, as discussed in Section~\ref{sec:hubbard}.

	To build up a tight-binding model for this topologically trivial band, we  construct localized Wannier states. For this purpose, we choose a gauge such that the bottom-layer component of the Bloch wave function  at each momentum is real and positive at the origin in real space. A linear superposition of such Bloch states leads to the Wannier state in Fig.~\ref{fig:wannier}, which is exponentially localized around the origin (one of the $\mathcal{R}_M^M$ sites) and threefold rotational symmetric. Appendix~\ref{appA} provides the detailed procedure to construct Wannier states.

\begin{figure}[t]
		\centering
		\includegraphics[width=0.6\columnwidth]{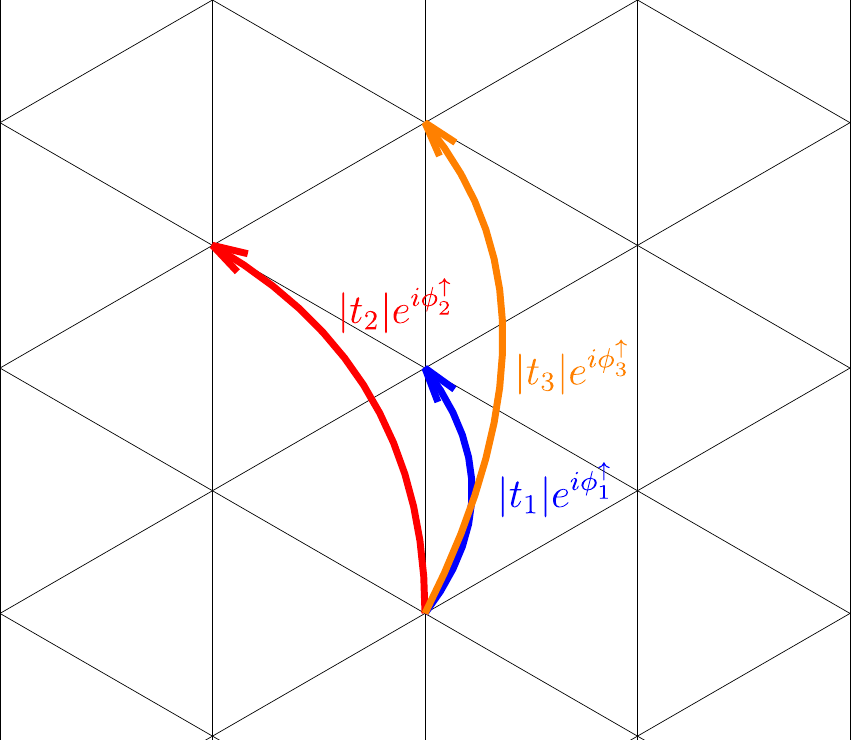}
		\caption{Illustration of representative hopping parameters in spin $\uparrow$ channel.}	
		\label{fig:lattice}
	\end{figure}

	The corresponding tight-binding model on the triangular lattice formed by $\mathcal{R}_M^M$ sites can be parametrized as
	\begin{equation}\label{eq:tb}
	H_{\text{TB}}=\sum_{s}\sum_{i,j}^{} t_{s}\left(\bm{R}_i-\bm{R}_j\right) c_{i,s}^\dagger c_{j,s},
	\end{equation}
	where $ s= \uparrow,\downarrow\ $ represents spin $\uparrow$ and $\downarrow$ states associated respectively with $ +K $ and $-K$ valleys, $ \bm{R}_i $ represents a site in the triangular lattice, and $ c_{j,s} $ ($ c_{j,s}^\dagger $) is  electron annihilation (creation) operator. $t_{s}\left(\bm{R}_i-\bm{R}_j\right)$ is the hopping parameter, which is constrained by the following relations. (1) Hermiticity of Hamiltonian~\eqref{eq:tb} requires that $  t_{s}\left(\bm{R}\right)= t_{s}^*\left(-\bm{R}\right)  $; (2) threefold rotational symmetry ($ C_3 $) requires that $  t_{s}\left(\bm{R}\right) =  t_{s}\left(\hat{\mathcal{R}}(2\pi/3)\bm{R}\right)  $, where $ \hat{\mathcal{R}}(2\pi/3) $ is a $2\pi/3$ rotation matrix; (3) time-reversal symmetry ($ \mathcal{T} $) requires that $  t_{s}\left(\bm{R}\right)= t_{-s}^*\left(\bm{R}\right) $. In Fig.~\ref{fig:lattice}, we use $ \abs{t_n} $ and $ \phi_{n}^{\uparrow} $ to denote the magnitude and phase for {\it representative} hopping parameters between $n$-th nearest neighbors in the spin $\uparrow$ channel. Since all the hopping terms within the $n$-th hopping shell are related by the aforementioned three relations, they can be determined once $ \abs{t_n} $ and $ \phi_{n}^{\uparrow} $ are determined.

	In Fig.~\ref{fig:hopping}, we present numerical values of $\abs{t_n}$ and $\phi_{n}^{\uparrow}$  for $n$ up to 3. Figure~\ref{fig:hopping}(a) shows that $ \abs{t_n} $ decays exponentially as the moir\'e periodicity increases (equivalently, the twist angle $ \theta $ decreases), since the Wannier states at different sites become further apart.  $\abs{t_n}$ and $\phi_{n}^{\uparrow}$ can also be controlled by $V_z$, as illustrated in Figs.~\ref{fig:hopping}(c) and \ref{fig:hopping}(d). An important effect is that the phase $\phi_{1}^{\uparrow}$ can be drastically changed by $V_z$.   $\phi_{1}^{\uparrow}$ is $\pi$ at $V_z=0$, and evolves to $4\pi/3$ ($2\pi/3$) when $|V_z|$ becomes large enough so that the two layers in the system become effectively decoupled. The dependence of $\phi_{1}^{\uparrow}$ on $V_z$ follows the change in the band structure shown in Fig.~\ref{fig:2D_band}. When  the hopping parameters take complex values (i.e., $\phi_{n}^{\uparrow}$ deviates from $0$ or $\pi$), they become spin dependent, which leads to effective spin-orbit couplings in the tight-binding model. As a very common feature, moir\'e systems have valley dependent band structures \cite{bistritzer2011moire, po2018origin, zhang2019bridging}, which, in our case, lead to the spin-orbit coupling because of spin-valley locking.

	\section{Hubbard model}\label{sec:hubbard}
\begin{figure}[t]
		\centering
		\includegraphics[width=1\columnwidth]{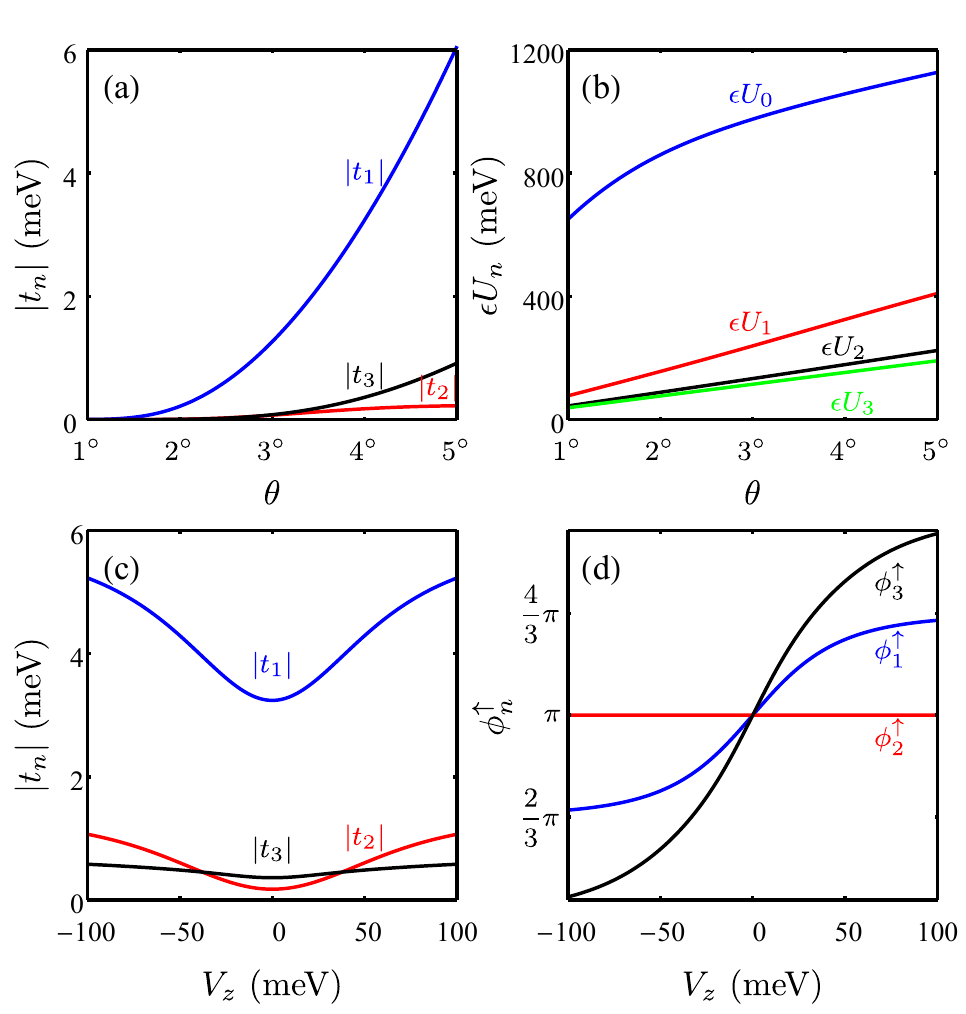}
		\caption{(a) $|t_n|$ and (b) $ \epsilon U_n$ as a function of the twist angle $ \theta $. $\epsilon$ is the effective dielectric constant. $V_z$ is 0 in (a) and (b). (c) $|t_n|$ and (d) $\phi_n^{\uparrow}$ as a function of $V_z$. $\theta$ is $4^{\circ}$ in (c) and (d).}	
		\label{fig:hopping}
	\end{figure}

	Many-body interactions are effectively enhanced for electrons in the moir\'e band with a narrow bandwidth because of the strongly suppressed kinetic energy. By combining  the tight-binding Hamiltonian in Eq.~\eqref{eq:tb} with electron-electron Coulomb repulsion, we can construct a generalized Hubbard model:
	\begin{eqnarray}\label{eq:hubbard}
	H&=&\sum_{s}\sum_{i,j}^{} t_{s}\left(\bm{R}_i-\bm{R}_j\right) c_{i,s}^\dagger c_{j,s}\nonumber\\
	&+&\frac{1}{2}\sum_{s,s'}\sum_{i,j}U(\bm{R}_i-\bm{R}_j) c_{i,s}^\dagger c_{j,s'}^\dagger c_{j,s'} c_{i,s},
	\end{eqnarray}
	where the repulsion $ U(\bm{R}_i-\bm{R}_j) $ between sites $i$ and $j$ is calculated by projecting the Coulomb repulsion $\tilde{U}(\bm{r})=e^2/(\epsilon r)$ onto the Wannier states. Here $\epsilon$ is the effective background dielectric constant that can be controlled by the three-dimensional dielectric environment. We take $\epsilon$ as a free parameter in our theory since its precise value is tunable (and not always precisely known). Numerical values of $U_0$ (on site repulsion) and $U_n$ ($n=1,2,3$ for repulsion between $n$-th nearest neighbors) are presented in Fig.~\ref{fig:hopping}(b). For a typical value of $\epsilon$ about 10, the on-site interaction $U_0$ can be at least one order-of-magnitude greater than the hopping parameters for twist angle $\theta$ below $5^{\circ}$. Therefore, tWSe$_2$ provides a platform to simulate the generalized Hubbard model on a triangular lattice. Moreover, the hopping parameters can be {\it in situ} controlled by an external displacement field. The effective interacting model is a generalized Hubbard model since both interaction and hopping in Eq.~\eqref{eq:hubbard} are not necessarily restricted to being on-site or nearest-neighbor, respectively as the whole many-body Hamiltonian matrix of Eq.~\eqref{eq:hubbard} can be calculated from our moir\'e band calculations for a given $ \epsilon $.

	\begin{figure}[t]
		\centering
		\includegraphics[width=0.9\columnwidth]{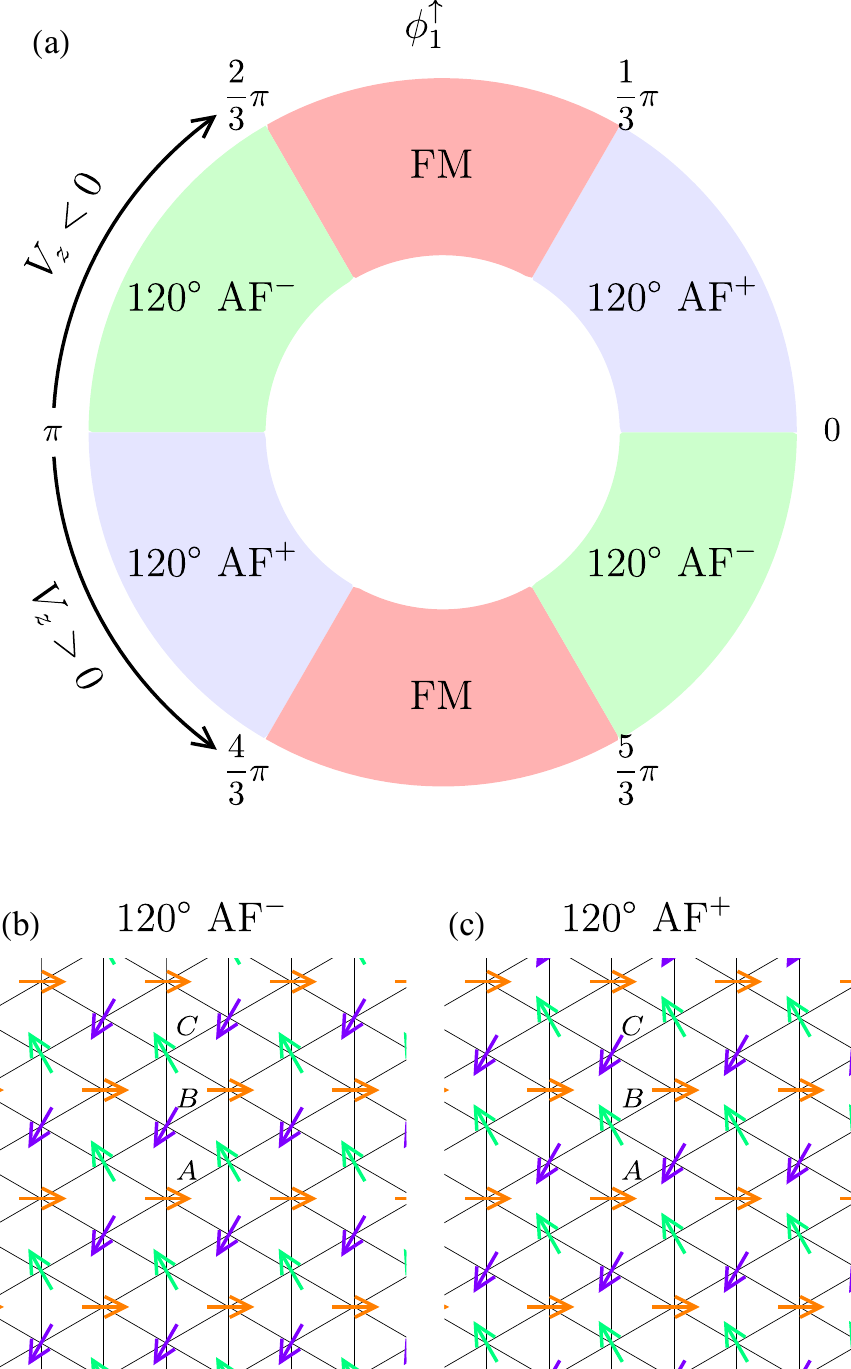}
		\caption{(a) Phase diagram of the Heisenberg model in Eq.~\eqref{eq:heisenberg} as a function of $\phi_1^{\uparrow}$. 120$^{\circ}$ AF$^{\pm}$ refer to in-plane antiferromagnetic phases shown in (b) and (c), while FM represents an in-plane ferromagnetic phase.   In tWSe$_2$, $\phi_1^{\uparrow}$ is constrained between $2\pi/3$ and $4\pi/3$, so only the 120$^{\circ}$ AF$^{\pm}$ phases are possible.}
		\label{fig:spin}
	\end{figure}

	\begin{figure}[t]
		\centering
		\includegraphics[width=1\columnwidth]{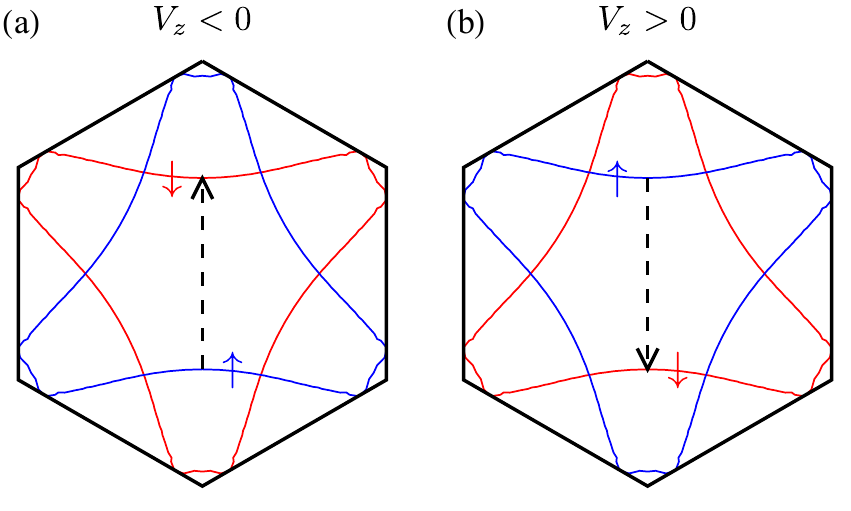}
		\caption{ Non-interacting Fermi surfaces at half filling for spin $\uparrow$ (blue) and $\downarrow$ (red). The dashed vectors indicate an approximate nesting between spin $\uparrow$  and $\downarrow$ Fermi surfaces. $V_z$ is negative in (a) and positive in (b).  }	
		\label{fig:fermi}
	\end{figure}

	\subsection{Heisenberg model}
	We consider carrier density at half-filling, where there is one electron per moir\'e unit cell in the first moir\'e valence bands  (equivalently, one hole per moir\'e unit cell when counting from the charge neutrality point of the twisted bilayer). The strong on-site repulsion $U_0$ suppresses double occupation at the same moir\'e site and gives rise to a Mott insulator. In this Mott limit (where $ U_0 $ is very large, much larger than the hopping parameters), the low-energy degrees of freedom are the electron spins at different sites. By retaining only nearest-neighbor hopping in the tight-binding model and on-site repulsion $U_0$, we can map the Hubbard model in Eq.~(\ref{eq:hubbard}) to spin Heisenberg model~\cite{macdonald1988fractu}:   
	\begin{equation}\label{eq:heisenberg}
		\begin{aligned}
		H= \frac{4\abs{t_1}^2}{U_0}\sum_{\langle i,j \rangle }^{\prime}  \Big( & S_{i}^z S_{j}^z+\cos (2\phi_{i,j}^{\uparrow})\sum_{\alpha=x,y}S_{i}^{\alpha} S_{j}^{\alpha} \\
		& +\sin (2\phi_{i,j}^{\uparrow}) \left(\bm{S}_{i}\times\bm{S}_{j}\right)\cdot \hat{z} \Big),
		\end{aligned}
		\end{equation}
	where the sum over $\langle i,j \rangle$ is restricted to nearest neighbors, the prime on the sum indicates that each pair of sites is counted only once, and $ \bm{S}_i $ is the spin-$ \frac{1}{2} $ operator at site $i$. Note that this mapping of the Hubbard model in Eq.~\eqref{eq:hubbard} to the Heisenberg model in Eq.~\eqref{eq:heisenberg} involves keeping only the on-site interaction and nearest-neighbor hopping as in the original minimal (rather than the generalized) Hubbard model.  In general, a more complete mapping of the full fermion model of Eq.~\eqref{eq:hubbard}, i.e., the generalized Hubbard model, to the spin model of Eq.~\eqref{eq:heisenberg} is, in principle, possible, but this involves complicated multi-spin terms beyond the Heisenberg model. This is unnecessary in the current problem since $ U_0 $ and $ t_1 $ indeed dominates the quantitative physics, thus allowing a mapping from an effective Hubbard model to an effective Heisenberg model of Eq.~\eqref{eq:heisenberg}. The first two terms in Eq.~\eqref{eq:heisenberg} are spin-exchange interactions as in a standard anisotropic Heisenberg model, while the last term $\left(\bm{S}_{i}\times\bm{S}_{j}\right)\cdot \hat{z}$ describes as an effective Dzyaloshinskii-Moriya (DM) interaction that is generated by the spin-orbit coupling inherent in the tight-binding model of Eq.~\eqref{eq:tb}. In Eq.~\eqref{eq:heisenberg}, $ \phi_{i,j}^s$ is the phase of the hopping parameter $t_{s}\left(\bm{R}_i-\bm{R}_j\right)$ between nearest-neighbor sites, and $ \hat{z} $ is the unit vector along out-of-plane direction. The relation $ \phi_{i,j}^{\downarrow} = -\phi_{i,j}^{\uparrow}$ is used in the simplification that leads to Eq.~\eqref{eq:heisenberg}.  One of the nearest-neighbor hopping phases in the spin-up channel is shown in Fig.~\ref{fig:hopping} as $ \phi_{1}^{\uparrow} $, and phases for other nearest-neighbor hopping parameters are related to $ \phi_{1}^{\uparrow} $ by the three relations given below Eq.~\eqref{eq:tb}. Therefore, the single parameter $ \phi_{1}^{\uparrow} $, which is tunable by the layer potential difference $V_z$, determines the ground state of the spin effective Heisenberg model in Eq.~\eqref{eq:heisenberg}.
	
	For $V_z=0$, $ \phi_{1}^{\uparrow} $ is $\pi$,  the DM interaction vanishes, and the model in Eq.~\eqref{eq:heisenberg} becomes the standard Heisenberg model with spin SU(2) symmetry on a triangular lattice. This isotropic Heisenberg model with only nearest-neighbor exchange coupling has a family of degenerate ground states with  three-sublattice $120^{\circ}$ long-range antiferromagnetic (AF) order, which we refer to as $120^{\circ}$ AF states.
	
	For a finite $V_z$, $ \phi_{1}^{\uparrow} $ deviates from $\pi$, and the finite DM interaction in the spin model (\ref{eq:heisenberg}) reduces the spin SU(2) symmetry down to U(1) symmetry, which originates from the valley U(1) symmetry in the Hubbard model of Eq.~\eqref{eq:hubbard}. In Fig.~\ref{fig:spin}, we show the calculated classical \textit{magnetic} phase diagram of Eq.~(\ref{eq:heisenberg}) as a function of $\phi_{1}^{\uparrow}$. This diagram is obtained by approximating the spin operator $\bm{S}_i$ as a classical vector with a fixed length and minimizing the energy with Luttinger-Tisza method~\cite{luttinger1946theory}. In our system, $ \phi_{1}^{\uparrow} $ takes values between $2\pi/3$ and $4\pi/3$, and crosses $ \pi $ when $ V_z $ crosses 0, resulting in a sign change in the DM interaction. The DM interaction acts as an anisotropy that favors in-plane spin ordering, and selects a subset of the $120^{\circ}$ AF states to be the ground state.  In particular, the spin ground states for $ V_z<0 $ and $ V_z>0 $ are, respectively, the $120^{\circ}$ AF$^-$ and $120^{\circ}$ AF$^+$ phases, which are demonstrated in Figs.~\ref{fig:spin}(b) and \ref{fig:spin}(c). To distinguish these two phases, we choose three sites ($A$, $B$, $C$) along a vertical line in the triangular lattice, as marked in Fig.~\ref{fig:spin}. The spins along the path $A \rightarrow B \rightarrow C$ rotate clockwise (anticlockwise) in the  $120^{\circ}$ AF$^-$ (AF$^+$) phase. Therefore, these two phases have opposite vector-spin-chirality orders that can be characterized by $\bm{S}_A \times \bm{S}_B$.

	\subsection{Mean-field theory}
	We also perform a Hartree-Fock mean-field study of the generalized Hubbard model defined by Eq.~\eqref{eq:hubbard} at half-filling. The mean-field calculation is not subject to the limit that $U_0 \gg |t_1|$ and provides an estimation of the charge excitation gap for the interaction-driven correlated insulator at half filling. We use the mean-field Ansatz from the spin configuration in the ground state of the Heisenberg model of Eq.~\eqref{eq:heisenberg}, i.e., the $120^{\circ}$ AF$^{+}$ (AF$^{-}$) states for positive (negative) $V_z$. These two different Ans\"atze for $V_z>0$ and $V_z<0$ can also be understood from Fermi surface instability. As shown in Fig.~\ref{fig:fermi}, the spin-$\uparrow$ and -$\downarrow$ Fermi surfaces in the non-interacting limit have an approximate nesting, with the opposite nesting vector in the momentum space for opposite signed $ V_z $. This approximate nesting can lead to interaction-driven instability in the spin-density-wave (SDW) channel.  The SDW order parameter can be taken as $\expval{c_{\bm{k}+ \bm{Q},\downarrow}^\dagger c_{\bm{k},\uparrow}}$ and $\expval{c_{\bm{k}- \bm{Q},\downarrow}^\dagger c_{\bm{k},\uparrow}}$, respectively, for $V_z>0$ and $V_z<0$. Here, we approximate  $\bm{Q}$ by the commensurate  wave vector $\bm{\kappa}_+-\bm{\kappa}_-$ that connects the two corners of the moir\'e Brillouin zone. The spin ordering wave vectors $\pm \bm{Q}$ are opposite for opposite signed $V_z$, following the Fermi-surface configurations shown in Fig.~\ref{fig:fermi}. The order parameters $\expval{c_{\bm{k} \pm \bm{Q},\downarrow}^\dagger c_{\bm{k},\uparrow}}$  in momentum space correspond to the  $120^{\circ}$ AF$^{\pm}$ state in real space. Therefore, the Heisenberg model in the strong coupling limit and the Fermi surface instability in the weak coupling limit are consistent with each other. 
	
	\begin{figure}[t]
		\centering
		\includegraphics[width=1\columnwidth]{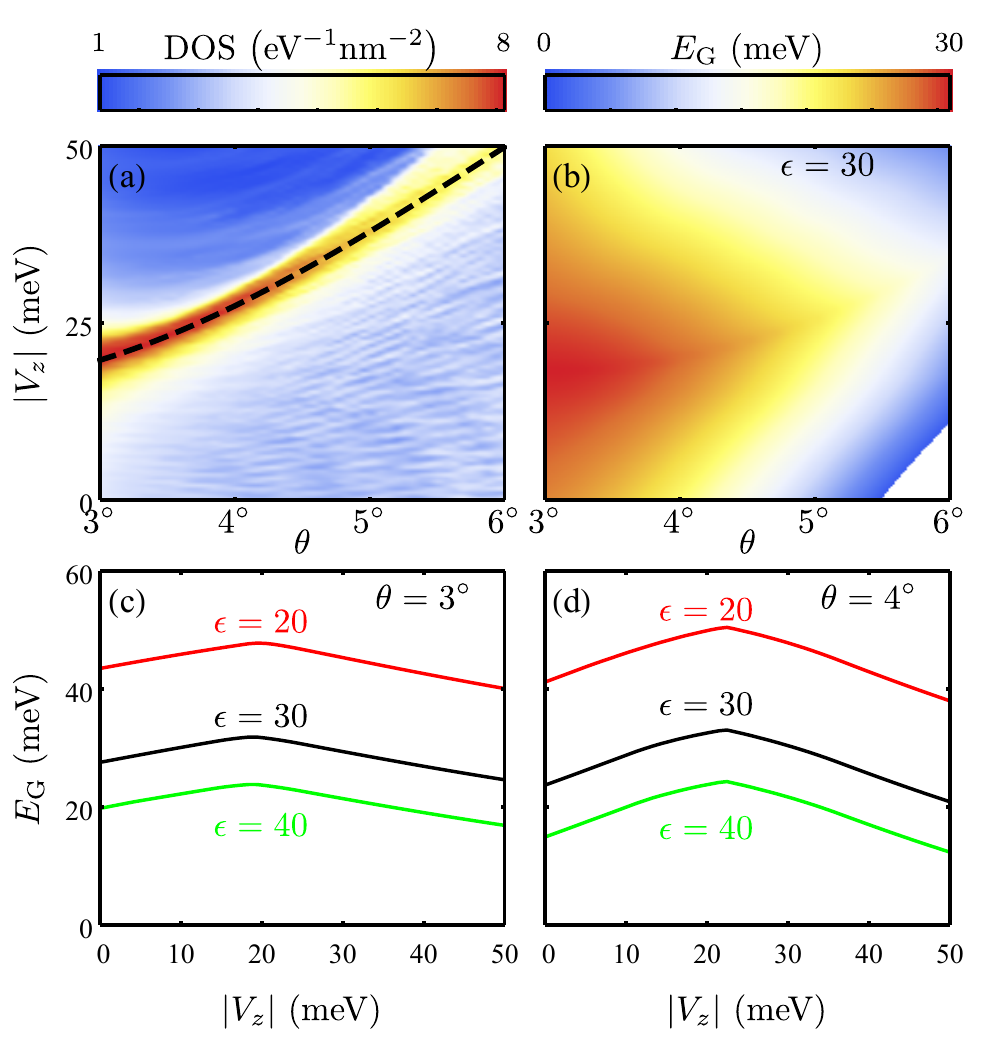}
		\caption{(a) The non-interacting density of states at half filling as a function of $\theta$ and $\abs{V_z}$. On the dashed line, van Hove singularities are at half filling. (b) The interaction-driven insulating gap as a function of  $\theta$ and $\abs{V_z}$. At a given $\theta$, the gap has a dome shape dependence on $\abs{V_z}$, which is illustrated in (c) and (d).}	
		\label{fig:mean_field}
	\end{figure}		
	
	With the above mean-field Ans\"atze, we perform a self-consistent mean-field calculation for the Hubbard model \eqref{eq:hubbard} that takes into account hopping up to the third nearest neighbors and the on-site Coulomb repulsion $U_0$. Including off-site Coulomb repulsion, which is much smaller than the on-site term $ U_0 $, where $ i $ and $ j $ are different in the second term of the right-hand side of Eq.~\eqref{eq:hubbard} is straightforward, but does not lead to any qualitatively different results (see Appendix \ref{appB}), essentially implying a small renormalization of the value of the Hubbard interaction $ U_0 $. The calculated charge gap $E_\text{G}$ at half-filling is shown in Fig.~\ref{fig:mean_field}. $E_\text{G}$ is finite for a large range of twist angle $\theta$. Therefore, there is no need to fine tune $\theta$ in tWSe$_2$ in order to realize correlated insulators as is absolutely necessary for twisted bilayer graphene, where the correlated insulator phase is very fragile. We find that $E_\text{G}$ has a strong dependence on $V_z$ particularly for weak interactions (large dielectric constant $\epsilon$). This follows the strong dependence of the non-interacting DOS as well as the nesting degree of Fermi surfaces at half filling on $V_z$, as demonstrated in Fig.~\ref{fig:2D_band} and also in Fig.~\ref{fig:mean_field}(a). A larger non-interacting DOS and a better nesting degree at half-filling implies a stronger interaction-driven instability towards symmetry-breaking states. As a result,  $E_\text{G}$ can have a dome-shape dependence on $V_z$, and the interaction-driven insulator at half-filling can be turned on and off by $V_z$ [Figs.~\ref{fig:mean_field}(b)-\ref{fig:mean_field}(d)]. We note that the calculated charge gap in Fig.~\ref{fig:mean_field} for reasonable values of $\epsilon$ are of the order of tens of meVs implying rather robust correlated insulating phases in twisted WSe$ _2 $.

	\begin{figure}[t]
		\centering
		\includegraphics[width=1\columnwidth]{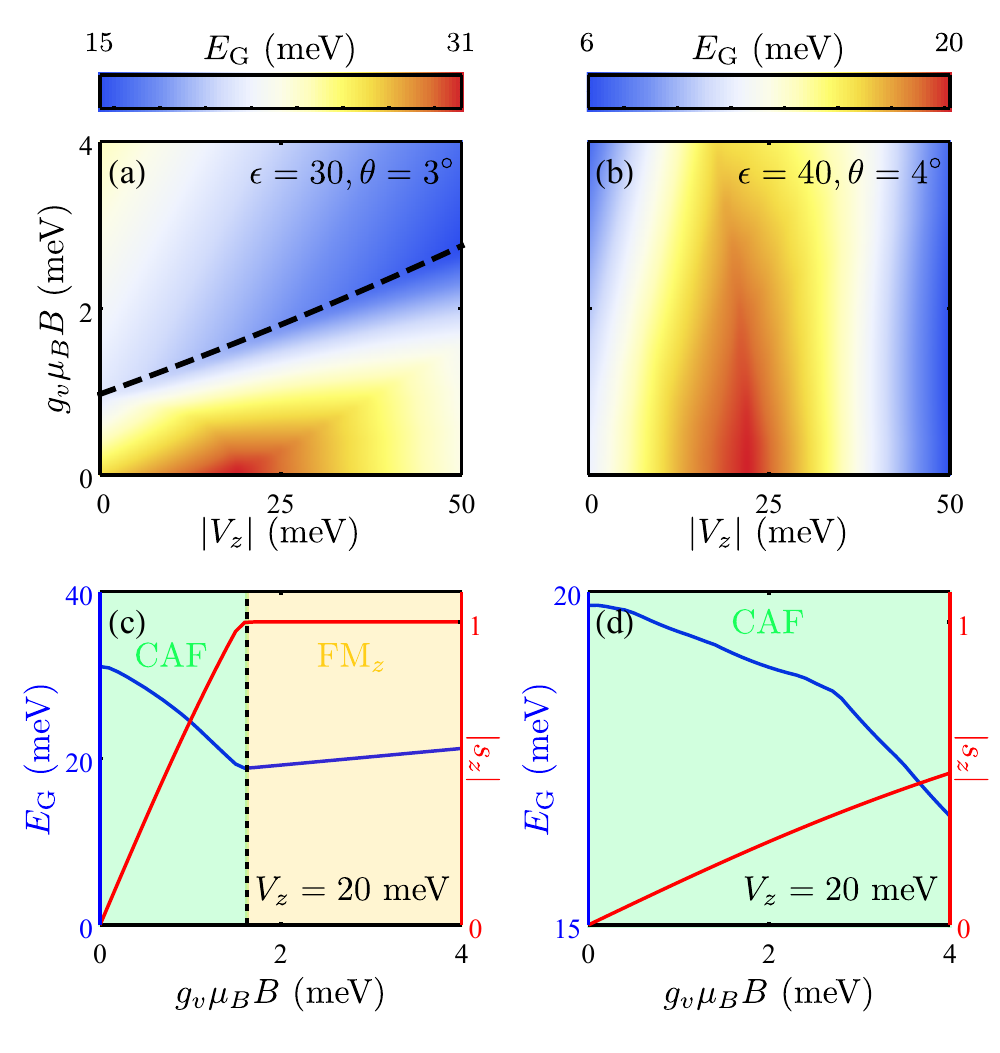}
		\caption{(a) The charge gap $E_\text{G}$ at half filling as a function $V_z$ and $B$. The dashed line marks $B_c$, above (below) which $E_\text{G}$ increases (decreases) with increasing $B$ field. (b) Similar plot as (a) but for a different set of parameter values. (c) and (d) are line-cut plots for (a) and (b), respectively. The red lines in (c) and (d) plot $s_z$, which is the spin polarization (per site) along $\hat{z}$ direction. $|s_z|=1$ represents full spin polarization.}	
		\label{fig:Zeeman}
	\end{figure}

	We further study the effect of an out-of-plane magnetic field $B \hat{z}$ on the half-filled correlated insulator, and consider the following Zeeman term:
	\begin{equation} \label{HZ}
	    H_{\text{Z}}=g_v \mu_B B \sum_{i} (c_{i,\uparrow}^{\dagger}c_{i,\uparrow}-c_{i,\downarrow}^{\dagger}c_{i,\downarrow})/2,
	\end{equation}
	where $\mu_B$ is the Bohr magneton. The effective $g$ factor $g_v$ has three contributions, $g_v=g_s+g_a+g_b$, where $g_s$, $g_a$, and $g_b$, respectively, capture the spin, atomic orbital, and Bloch band contributions. The spin $g$ factor $g_s$ is 2. The atomic orbital characters of states at the $\pm K$ valley valence band maximum are mainly $d_{x^2-y^2} \pm i d_{xy}$, and the atomic $g$ factor is therefore $g_a=2-(-2)=4$. Electrons in Bloch bands carry an additional orbital magnetic moment \cite{xiao2007valleycontrasting}, which contributes to $g_b$. Since $g_b$ depends on the details of the moir\'e band structure, we do not present a quantitative estimation of it. In the Mott limit where electrons are strongly localized, $g_b$ should only lead to a small correction. We take  $g_v$ as a phenomenological parameter, and expect it to be of the same order of magnitude as $g_s+g_a=6$.
	
	In the presence of the $B \hat{z}$ field, the 120$^\circ$ AF$^{\pm}$ states turn into canted antiferromagnets (CAF). We calculate the charge gap $E_\text{G}$ at half-filling as a function of $B$, and show representative results in Fig.~\ref{fig:Zeeman}. With other parameters fixed, there is a critical field $B_c$. For $|B|<B_c$, the ground state is in the CAF phase with spins canted towards the out-of-plane direction, and $E_\text{G}$ decreases with increasing $|B|$ field due to a loss of exchange energy. For $|B|>B_c$, the ground state is in a ferromagnetic state (FM$_z$) with all spins aligned in the out-of-plane direction, and  $E_\text{G}$ increases with increasing $|B|$ field due to the Zeeman energy. The FM$_z$ phase is also a valley-polarized state. This valley-polarized state carries zero (finite) Chern number when the first moir\'e valence bands are in the topologically trivial (nontrivial) phase, and supports vanishing (quantized) anomalous Hall effect, which provides a mechanism to identify the band topology using transport measurement. Here, we focus on the topologically trivial moir\'e bands, and the corresponding FM$_z$ phase has no anomalous Hall effect.
	
	 We can also estimate the critical field $B_c$ from the Heisenberg model in Eq.~(\ref{eq:heisenberg}), and $B_c$ estimated in this way is proportional to the spin-exchange interaction $J_1=4|t_1|^2/U_0$. Therefore, $B_c$ is expected to decrease with decreasing twist angle $\theta$, following the weakening of $J_1$ at smaller $\theta$. The mean-field results shown in Fig.~\ref{fig:Zeeman} are indeed consistent with this $\theta$ dependence of $B_c$.

	\subsection{Comparison with transport experiment}
    We compare our theoretical studies with the transport experiment on tWSe$_2$ in Ref.~\onlinecite{wang2019magic}. This experimental paper \cite{wang2019magic} presents transport study on multiple devices of tWSe$_2$ with the twist angle $\theta$ in the range between $4^{\circ}$ and $5^{\circ}$, and reports correlated insulators at half-filling of the first moir\'e valence bands. Our theory is consistent with this experiment \cite{wang2019magic} in key aspects as discussed in the following. 
    
    The measured van Hove singularities determined from Hall effect have a strong dependence on displacement field~\cite{wang2019magic}. This behavior is captured by our band structure calculation shown in Fig.~\ref{fig:2D_band}, which shows that the van Hove singularities can be tuned from below to above half filling by $V_z$. 
    
    The correlated insulators at half-filling develop for a large range of twist angle up to about $5^{\circ}$, and is controllable via displacement field~\cite{wang2019magic}. Our mean-field calculation shown in Fig.~\ref{fig:mean_field} provides a qualitative description of this observation.  In particular, we also find a dome-shape dependence of the insulating gap at half-filling on the layer potential difference, as in the experiment~\cite{wang2019magic}. This dome-shaped experimental insulating gap is a few ($ \sim $2-4) meV typically in Ref.~\onlinecite{wang2019magic} rather than being $ > $10-40 meV or so as we find mostly for the excitation gap in our theory. We note, however, that the measured insulating gap in Ref.~\onlinecite{wang2019magic} is even quantitatively consistent with our theoretical charge gap in Fig.~\ref{fig:mean_field} for a large value of $ \epsilon $ (40 or above).  This quantitative agreement for large dielectric constant should not be taken too seriously because our mean-field theory is bound to overestimate the magnitude of the gap and the experiment measures a transport activation gap which is typically much smaller than the theoretical excitation gap.

    The correlated insulating gap at half-filling is experimentally found to decrease with increasing out-of-plane magnetic field  when the field is weak \cite{wang2019magic}. Our theoretical results shown in Fig.~\ref{fig:Zeeman} agree with this observation for weak $B$ fields. When the $B$ field is strong enough, it can drive a spin- (equivalent to valley-) polarized insulating state, of which the charge gap becomes an increasing function of $B$. Therefore, the charge gap at half-filling can have a nonmonotonic dependence on $B$, which has also been experimentally observed \cite{wang2020private}. The absence or presence of anomalous Hall effect in the spin- (valley-) polarized insulator at half-filling provides a transport signature to determine the topological nature of the moir\'e bands.

	\section{Conclusion}\label{sec:conclusion}
	In summary, we present a systematic theoretical study of tWSe$_2$, and demonstrate the perspective of using this moir\'e system as a platform to realize interesting single-particle physics as well as many-body physics. For the single-particle moir\'e bands, we calculate the topological phase diagrams characterized by the valley-contrast Chern numbers. By comparing the theoretical LDOS with STM measurements~\cite{zhang2020flat}, we conclude that the first moir\'e valence band is likely to be topologically trivial, whereas the second and third moir\'e valence bands are likely to be topological. By increasing the hole density in the system, it should be possible to study the topological moir\'e bands experimentally if one can push the Fermi level into the higher moir\'e bands.
	
	For the interacting physics, we focus on the first moir\'e band, and construct a generalized Hubbard model. We show that tWSe$_2$ can act as a highly tunable Hubbard model simulator. In particular, the layer potential difference $V_z$ can drastically change the non-interacting moir\'e bands, control the charge excitation gap of the correlated insulators at half-filling, and generate an effective DM interaction in the corresponding spin Heisenberg model at half-filling. The moir\'e bands in tWSe$_2$ are relatively flat over a large range of twist angles $\theta$. Therefore, observation of correlation effects does not require fine tuning of $\theta$ in this system, which represents an advantage compared to TBG. 
	
	We envision that several directions can be explored following our theory. The transport experiment in Ref.~\cite{wang2019magic} has been limited to filling factors within the first moir\'e valence bands, which are likely to be topologically trivial. It would be interesting to increase the hole-doping level, and perform transport study in the second and even third moir\'e valence bands which likely carry finite valley-contrast Chern numbers. The spin-dependent Berry curvatures in these bands can lead to large spin Hall effect. The  enhanced Coulomb interactions may drive valley polarization, which, combined with the finite valley Chern number, can lead to quantum anomalous Hall effects. 
	
	We show that a field-tunable DM interaction can be realized in the spin Heisenberg model. This DM interaction pins vector spin chirality of the antiferromagnetic ground state. It is desirable to explore effects of DM interaction on spin and magnon transport, and find experimental probes that can distinguish opposite  vector spin chiralities. The 120$^{\circ}$ AF$^{\pm}$ states spontaneously break the U(1) symmetry of the Heisenberg model in Eq.~\eqref{eq:heisenberg}, which can then support spin superfluidity.
	
	We construct a generalized Hubbard model on triangular lattice with a field-tunable spin-orbit coupling, and study this model at half-filling by mapping it to the Heisenberg model as well as using a mean-field theory. It is conceivable to investigate this model using other techniques and also at other filling factors. The Hubbard model on triangular lattice can potentially host a variety of intriguing phases, for example,  quantum anomalous Hall insulators \cite{martin2008itinerant}, chiral superconductors \cite{nandkishore2014superconductivity},  and even spin liquids \cite{szasz2020chiral}. The inclusion of spin-orbit coupling should enrich the physics. Possible signatures of superconductivity in tWSe$_2$ have been reported in Ref.~\cite{wang2019magic}.  Our theoretical model can be a starting point to address exotic many-body physics including superconductivity in this system.
	
	While our theory focuses on ground-state physics, collective excitations, for example, excitons, in moir\'e pattern can also be very interesting \cite{wu2017topological,wu2018theory,tran2019evidence}. The realization of correlated insulators in tWSe$_2$ combined with the strong light-matter interaction already present in TMDs opens up the possibility to study optical physics in the strongly correlated regime.
	
    \section{Acknowledgment}
    F. W. thanks L. Wang, E.-M. Shih, A. Ghiotto and B. LeRoy for valuable discussions and sharing unpublished data. This work is supported by the Laboratory for Physical Sciences.

	\begin{figure}[b]
		\includegraphics[width=\columnwidth]{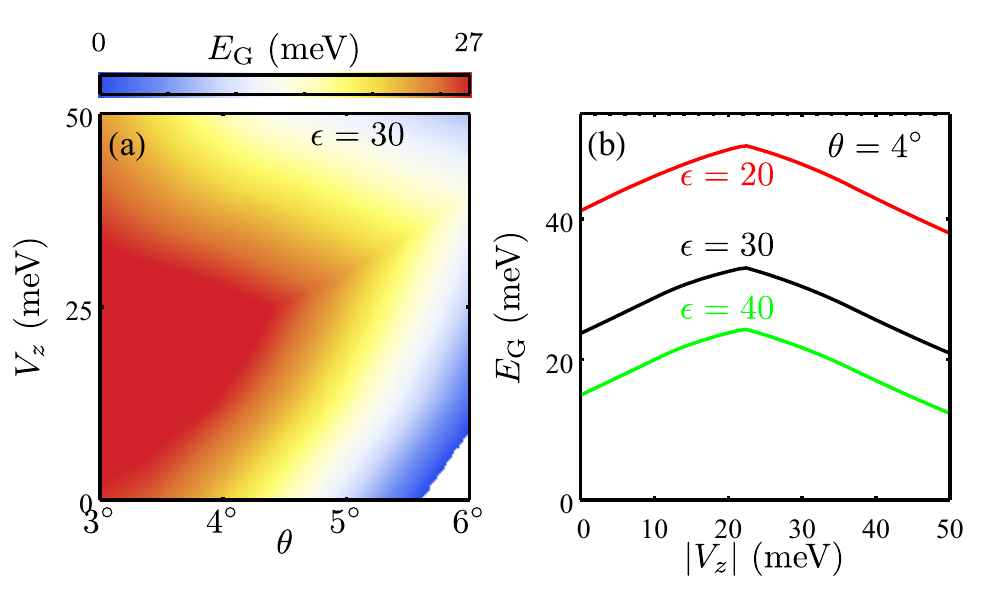}
		\caption{(a) The interaction-driven insulating gap $E_\text{G}$ at half filling as a function of $ \theta $ and $ \abs{V_z} $. The mean-field calculation is done by including interaction terms with $( U_0, U_1, U_2, U_3 )$. (b) A line-cut plot of (a) at $ \theta=4^\circ $. The gap $E_\text{G}$ only differs from Fig.~\ref{fig:mean_field} quantitatively while the dome feature as a function of $ \abs{V_z} $ does not change.}
		\label{fig:mean_field_U3}
	\end{figure}
	
	\appendix
	\section{WANNIER STATES AND HUBBARD MODEL PARAMETERS}
	\label{appA}
	When the first moir\'e valence band is topologically trivial, the corresponding Wannier state $W(\bm{r})$ located at the site $\bm{R}=\bm{0}$ and associated with $+K$ valley can be constructed as follows
	\begin{equation}
	W(\bm{r})=\frac{1}{\sqrt{N}}\sum_{\bm{k}\in\text{BZ}}\psi_{\bm{k}}(\bm{r}),
	\end{equation}
	where the momentum $\bm{k}$ is summed over the first moir\'e Brillouin zone (BZ), and $N$ is the number of $\bm{k}$ points in the summation.  $\psi_{\bm{k}}(\bm{r})$ is the Bloch wave function of the moir\'e Hamiltonian $\mathcal{H}_\uparrow$ in Eq.~\eqref{eq:moire}  and can be represented by a two-component spinor $[\psi_{\bm{k},b}(\bm{r}), \psi_{\bm{k},t}(\bm{r})]$ in the layer pseudospin space. Correspondingly, $W(\bm{r})$  is also a two-component spinor. We choose the phase of $\psi_{\bm{k}}(\bm{r})$ such that its bottom layer component is real and positive at the origin in real space, namely, $\psi_{\bm{k},b}(\bm{r}=\bm{0}) > 0$ for every $\bm{k}$. With this gauge, we obtain a symmetric Wannier state located at $\bm{R}=\bm{0}$, as illustrated in Fig.~\ref{fig:wannier}. Wannier states located at a generic lattice site $\bm{R}$ are obtained through lattice translation, $W_{\bm{R}}(\bm{r})=W(\bm{r}-\bm{R})$.
	
	The hopping integral in the tight-binding model is calculated by
	\begin{eqnarray}
	t_\uparrow(\bm{R}_i-\bm{R}_j)&=&\int W_{\bm{R}_i}^*(\bm{r}) \mathcal{H}_\uparrow W_{\bm{R}_j}(\bm{r}) d^2\bm{r}\nonumber\\
	&=&\frac{1}{N}\sum_{\bm{k}\in\text{BZ}}e^{i\bm{k}\cdot\qty(\bm{R}_i-\bm{R}_j)}\varepsilon_{\bm{k},\uparrow},
	\end{eqnarray}
	where $ \varepsilon_{\bm{k},\uparrow} $ is the band energy of the first moir\'e valence band in $+K$ valley. We find that $ \varepsilon_{\bm{k},\uparrow} $ can be accurately reconstructed by including hoppings up to the third nearest neighbors in the tight-binding model.  By  time-reversal symmetry, the  Wannier state at site $\bm{R}$ associated with  $-K$ valley can be defined to be $ W_{\bm{R}}^*(\bm{r})$ .

	The density-density Coulomb interaction $ U $ between two sites $ \bm{R}_i$ and  $ \bm{R}_j $ is given by
	\begin{equation}\label{eq:U}
	\begin{aligned}
	  &U(\bm{R}_i-\bm{R}_j) \\
	=& \int d^2 \bm{r}_1 d^2 \bm{r}_2 V(\bm{r}_1-\bm{r}_2) \abs{W_{\bm{R}_i}(\bm{r}_1)}^2\abs{W_{\bm{R}_j}(\bm{r}_2)}^2 \\
	=& \int \frac{d^2 \bm{q}}{\qty(2\pi)^2} V(\bm{q}) |M(\bm{q})|^2  e^{i\bm{q}\cdot\qty(\bm{R}_i-\bm{R}_j)} ,
	\end{aligned}
	\end{equation}
	where $ V(\bm{r})=e^2/(\epsilon \abs{\bm{r}}) $ is the Coulomb interaction, $V(\bm{q})=2\pi e^2/(\epsilon\abs{\bm{q}})$ is its Fourier transform, and $M(\bm{q})$ is defined by
	\begin{equation}
	M(\bm{q})=\int d^2\bm{r} \abs{W(\bm{r})}^2 e^{i\bm{q}\cdot \bm{r} }.
	\end{equation}
	
	We take the dielectric constant $\epsilon$ to be a constant that is determined by the environmental screening. This is an approximation that neglects the frequency and position dependence of $\epsilon$. Higher-energy moir\'e bands, which are neglected in the construction of the interacting model, can generate frequency dependent interactions (equivalently, $\epsilon$ becomes frequency dependent) \cite{Aryasetiawan2004}. In addition, environmental screening from the encapsulating material can be highly nonlocal, which effectively makes $\epsilon$ to be position dependent \cite{Cho2018}. We expect that these complication do not change our qualitative results, and leave them to future study.

	\section{MEAN-FIELD RESULTS WITH REMOTE INTERACTIONS}
	\label{appB}
	We perform the mean-field calculation of the Hubbard model by performing Hartree-Fock decomposition of the interaction terms, following procedures discussed in Ref.~\onlinecite{Guiliani2005}. The mean-field equation is solved through iterations, with initial ansatz from the  magnetic phase of the Heisenberg model. We have also used random spin configurations within a $\sqrt{3}\times \sqrt{3}$ magnetic supercell as initial inputs, and found that the final self-consistent mean-field solution does not change.
	
	In the main text,  mean-field results shown in Fig.~\ref{fig:mean_field} are obtained by including only the on-site Coulomb interaction $ U_0 $ in the Hamiltonian. We have also performed the mean-field calculation by taking into account remote interactions up to $U_3$; the corresponding results presented in Fig.~\ref{fig:mean_field_U3} demonstrate that interactions beyond the onsite repulsion $U_0$ do not lead to a qualitative change of the correlated insulating gap at half filling. This provides a justification on why we can only consider the onsite repulsion when studying interaction effects at half filling. Physically, the correlated insulating state at half filling has the nature of a Mott insulator, and is driven primarily by the onsite repulsion $U_0$.

	\bibliographystyle{apsrev4-1}
	\bibliography{TMD}

\end{document}